%% file: main_manuscript_imwut.tex
\documentclass[acmlarge, nonacm, screen, table, xcdraw]{acmart}
\usepackage{algpseudocode}
\usepackage{algorithm}
\usepackage{amsmath}
\usepackage{lipsum}
\usepackage[utf8]{inputenc} 
\usepackage[T1]{fontenc}    

\usepackage{hyperref}
\usepackage{booktabs}       
\usepackage{amsfonts}       
\usepackage{nicefrac}       
\usepackage{microtype}      
\usepackage{xcolor}         
\usepackage{wrapfig}
\usepackage{subcaption}
\usepackage{tabularx}
\usepackage{booktabs}
\usepackage{microtype}
\usepackage{graphicx}
\usepackage{booktabs} 
\usepackage{amsmath}
\usepackage{mathtools}
\usepackage{amsthm}
\usepackage{mathrsfs}
\usepackage{graphicx}

\usepackage[normalem]{ulem}
\usepackage{colortbl}
\usepackage{capt-of}
\usepackage[]{graphicx}
\usepackage{multirow}
\usepackage{algorithm}
\usepackage{algpseudocode}
\usepackage{amsmath}
 \usepackage{graphicx}
\usepackage{pifont}
\usepackage{graphicx}
\usepackage{caption}
\usepackage[capitalize,noabbrev]{cleveref}
\usepackage{gensymb}

\newcommand{\locnet}{\texttt{HALo}}
\newcommand{\clfnet}{\texttt{CoCo}}
\newcommand{\jointnet}{\texttt{HALo-CoCo}}

\usepackage[utf8]{inputenc} 
\usepackage[T1]{fontenc}    
\usepackage{url}            
\usepackage{booktabs}       
\usepackage{amsfonts}       
\usepackage{nicefrac}       
\usepackage{microtype}      
\usepackage{xcolor}         
\usepackage{tcolorbox} 
\tcbuselibrary{skins}
\definecolor{Orange50}{HTML}{FFF3E0}
\newtcolorbox[auto counter]{Formulation}[2][]{%
  colback=Orange50, 
  colframe=Orange50!85!black, 
  fonttitle=\bfseries, 
  title=Task Formulation~\thetcbcounter: #2, 
  coltitle=black, 
  boxrule=0.5pt, 
  arc=4pt, 
  left=1mm, 
  right=1mm, 
  top=1mm, 
  bottom=1mm, 
  enhanced, 
}
\usepackage{etoolbox}
\definecolor{brightblue}{HTML}{007BFF}
\newcommand{\imwut}[1]{{#1}}
\AtBeginDocument{%
  }

\setcopyright{acmlicensed}
\copyrightyear{2018}
\acmYear{2018}
\acmDOI{XXXXXXX.XXXXXXX}

\acmConference[Conference acronym 'XX]{Make sure to enter the correct
  conference title from your rights confirmation emai}{June 03--05,
  2018}{Woodstock, NY}
\acmISBN{978-1-4503-XXXX-X/18/06}

\begin{document}






\title{Towards Localizing Conversation Partners using Head Motion}



\author{Payal Mohapatra}
\authornote{Work conducted during internship at Meta Reality Labs, Redmond, USA }
\email{payal.mohapatra@nortwhestern.edu}
\affiliation{%
  \institution{Northwestern University}
  \city{Evanston}
  \state{Illinois}
  \country{USA}
}

\author{Calvin Murdock}
\affiliation{%
  \institution{Meta Reality Labs}
  \city{Redmond}
  \country{USA}
}
\author{Ali Aroudi}
\affiliation{%
  \institution{Meta Reality Labs}
  \city{Redmond}
  \country{USA}
  }

\author{Ishwarya Ananthabhotla}
\affiliation{%
  \institution{Meta Reality Labs}
  \city{Redmond}
  \country{USA}
  }
\author{Anjali Menon}
\affiliation{%
  \institution{Meta Reality Labs}
  \city{Redmond}
  \country{USA}
}
\author{Buye Xu}
\affiliation{%
  \institution{Meta Reality Labs}
  \city{Redmond}
  \country{USA}
  }

\author{Morteza Khaleghimeybodi}
\email{morteza@meta.com}
\affiliation{%
  \institution{Meta Reality Labs}
  \city{Redmond}
  \country{USA}
  }


\begin{abstract}
 
Many individuals struggle to understand their conversation partners in noisy settings, particularly amidst background speakers or due to hearing impairments. Under such circumstances, emerging wearables like smartglasses offer a transformative opportunity to enhance speech from conversation partners. Crucial to this is the ability to identify the direction in which the user is interested in listening, which we refer to as the user's acoustic zones of interest. While current spatial audio-based methods are effective in resolving the direction of vocal input, they are agnostic to the user's listening preferences and have limited functionality in noisy settings and when interfering speakers are present. To address this, there is a need to actively incorporate behavioral cues for inferring a user's acoustic zones of interest. Motivated by this need, we explore the effectiveness of the user’s head-orienting behavior, captured by the Inertial Measurement Units (IMUs) on smartglasses, as a modality for localizing acoustic zones of interest in seated conversations. We introduce \locnet{}, a head-orientation-based acoustic zone localization network that leverages smartglasses' IMUs to non-invasively infer auditory zones of interest corresponding to conversation partner locations. By integrating an \emph{a priori} estimate of the number of conversation partners, our approach demonstrates a 21\% performance improvement over existing methods. We complement this with \clfnet{}, which classifies the number of conversation partners using only IMU data, achieving an accuracy of 0.74; compared to rule-based and generic time-series models, this yields a 35\% gain in task performance. We discuss practical considerations for feature extraction and inference and provide qualitative analyses over extended sessions. \imwut{We also demonstrate a minimal end-to-end speech enhancement system and show that a head-orientation-based localization scheme can offer clear advantages in extremely noisy settings with multiple conversation partners.} Our work demonstrates the feasibility of a non-visual behavioral modality for inferring users’ conversational preferences with minimal sensing requirements, and highlights its potential to enhance conversational immersiveness through smartglasses.

\end{abstract}


\maketitle

\input{sections/introduction}

\input{sections/relatedworks}

\input{sections/methodology}

\input{sections/evaluation}

\input{sections/discussion}
\input{sections/conclusion}

\bibliographystyle{ACM-Reference-Format}
\bibliography{sample-base}

\appendix
\input{sections/appendix}

\end{document}

%% file: sections/introduction.tex
\section{INTRODUCTION}

\begin{figure}
    \centering
    \includegraphics[width=0.7\linewidth]{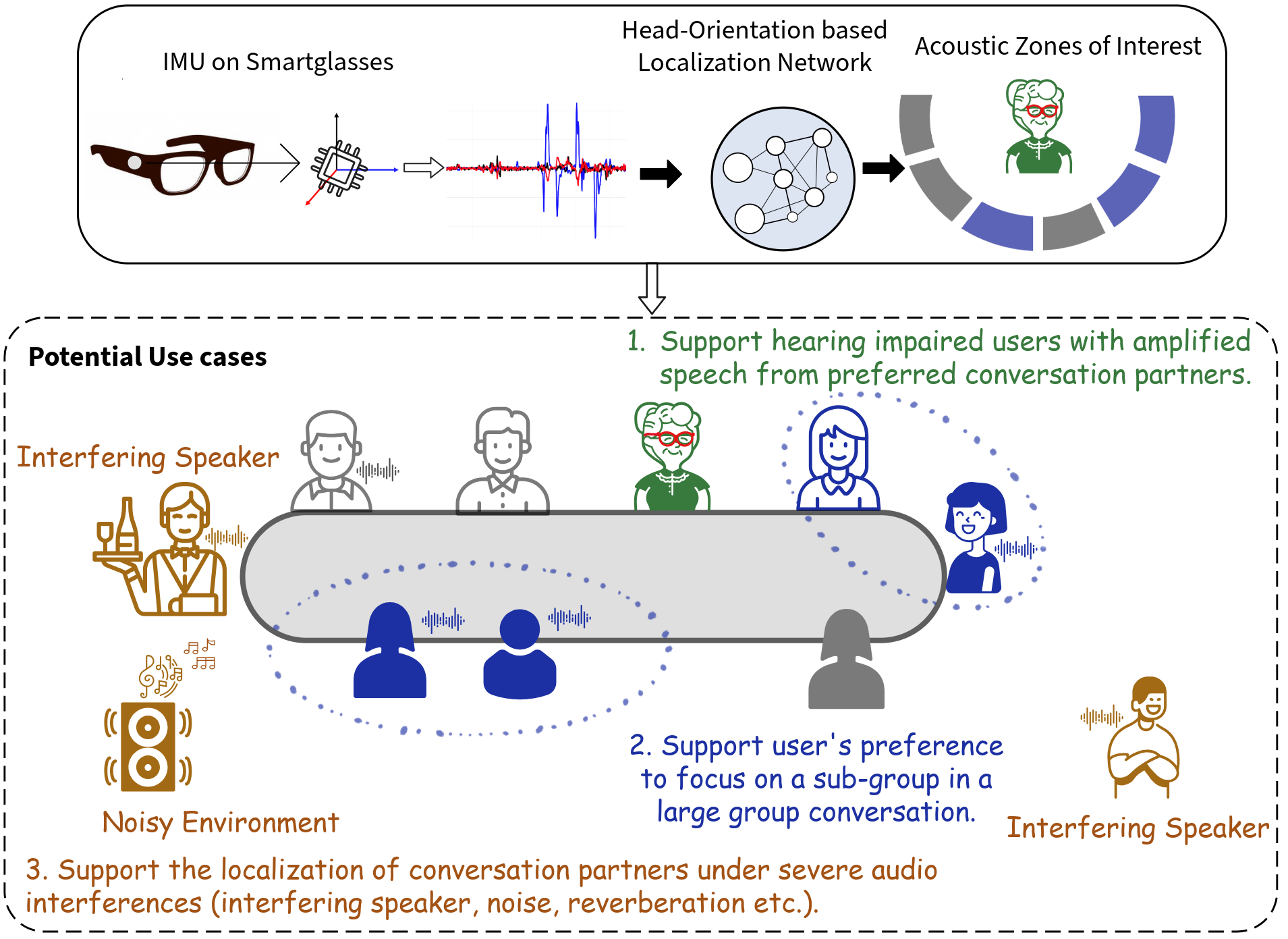}
    \vspace{-8pt}
    \captionsetup{font=small}
    \caption{\small \textbf{Concept}: Using IMU sensors on smartglasses to extract the head-orienting patterns of the user during a conversational setting to identify the user's acoustic zones of interest (which are indicative of the conversation partners' locations). \\
    \textbf{Potential use cases}: Consider a large multiparty noisy setting (e.g., a restaurant) with interfering/background speakers. The user (shown wearing smartglasses) engages with a preferred subgroup. These conversation partners are shown in \textcolor{blue}{blue}. A technology that leverages user behavior-driven localization of conversation partners as proposed in this work (1) can support hearing-impaired users through directional speech enhancement based on their listening preferences, (2) can support a user's intentional engagement with preferred conversation partners in a large conversation group, and (3) provides robust performance under noisy acoustics and in the presence of interfering speakers.
    }
    \vspace{-15pt}
    \label{fig:overall_motivation}
\end{figure}


Group conversations are central to daily social life, yet individuals with hearing impairments~\cite{strawbridge2000negative, gatehouse2004speech, mehra2020potential} or hearing fatigue in noisy environments~\cite{mckellin2007pragmatics, maruyama2020assessing} encounter significant barriers to effective communication. Traditional approaches to improving speech intelligibility typically enhance the signal-to-noise ratio (SNR) through beamforming, spatial filtering, or neural network-based speech enhancement techniques~\cite{moore2019personalized, van1988beamforming, chatterjee2022clearbuds, xu24i_interspeech}. The recent proliferation of smartglasses as a ubiquitous wearable platform presents an opportunity to redefine the conversational experience by deploying on-device beamforming or neural network-based speech enhancement algorithms~\cite{ruofei2024foveated, matsuo2024360, mehra2020potential}, which preserve signals from a target direction while attenuating those from other directions. Determining the target direction of interest—particularly accounting for the listener's preferences for conversation partners—is crucial for the success of such technology~\cite{chen2024hearable}. However, current audio-based modalities do not incorporate user preferences, may inadvertently amplify interfering speakers, and often exhibit limited functionality under noisy or non-ideal acoustic conditions~\cite{battula2025robust}. This highlights the need to explore alternative behavioral modalities that can implicitly infer listeners' preferences and guide conversation-enhancing algorithms~\cite{ruofei2024foveated, matsuo2024360}. Motivated by this practical need, in this work we investigate the head motion of the listener as a feasible modality to infer acoustic zones of interest using smartglasses, which approximate the locations of conversation partners. Acoustic zones of interest are defined based on a discretized representation of azimuth angles relative to the listener's average facing direction.


Identifying the spatial locations of conversation partners is a foundational task for enabling immersiveness in conversational settings using smartglasses. Prior works~\cite{friedland2009visual, wu2021binaural, busso2005smart} largely leverage spatial audio-based techniques (captured using microphone arrays) to determine speaker locations. While these methods are generally successful, they are prone to misidentifying interfering speakers, are susceptible to background noise, and fail to account for a listener's preferences for instance listener's preference towards a fixed subgroup of speakers in large multi-party conversations as illustrated in Figure~\ref{fig:overall_motivation}. This highlights the need to identify conversation partners specifically, rather than all speakers in the scene. Some proposed methods address these shortcomings by incorporating egocentric visual modalities for better contextualization~\cite{choi2006probabilistic, jiang2022egocentric}. While these methods offer superior performance for identifying and localizing conversation partners under noisy conditions, they can be intrusive in social settings and demand significant computational resources, making them infeasible for wearable platforms.


Beyond audio-visual modalities, recent works in speaker localization have shown the benefit of incorporating behavioral modalities like head-orientation and gaze either as a proxy for ground-truth annotations to enable more large-scale self-supervised training~\cite{murdock2024self} or as a mode of understanding the visual focus of attention using gaze~\cite{li2013learning,mondal2023gazeformer, sidenmark2019eye}. This motivates us to leverage existing on-device sensors to capture orienting behaviors and provide better context for user preferences. Parallely, extensive literature~\cite{tiede2019noggin,mcclave2000linguistic,mcgarrigle2021older,hendrikse2019movement,brimijoin2010auditory} in auditory neuroscience also highlight the role of head orientation as a useful modality for linguistic behaviors in controlled settings with manual annotations on small-scale datasets. Some works like ~\citep{lu2022sound} have extended these neuroscience results to consider behavioral modalities independently by formulating tasks like active speaker selection from a given set of speaker locations based on head orientations, or understanding how group dynamics can help in determining accurate head orientations~\cite{tan2021multimodal}. However, much of this orienting information is derived from egocentric or exocentric videos, which do not address the constraints of relying on visual modalities for wearable platforms. Some works have used self-voice-based head orientation detection using external microphones~\cite{yang2021model}, which may fail under multiple speaker scenarios in a typical conversation setting. Thus, the challenge of determining acoustic zones of interest independent of prior knowledge of the number of speakers and their original precise locations remains an unexplored but important setting. In this work, we investigate this challenging question: \textbf{\textit{What is the potential of using only head orientation information to determine a listener's acoustic zones of interest in natural conversations?}} 

We propose \imwut{a novel task of} leveraging commonly available sensors on smart glasses, such as IMUs, to detect conversation-focused orienting behaviors and infer the listener's acoustic zones of interest. An overview is illustrated in Figure~\ref{fig:overall_motivation}. However, there are several challenges in successfully leveraging head orientation to identify acoustic zones of interest. First, as a behavioral modality, it is inherently weak in explicitly capturing linguistic behaviors and is instead coupled with other behavioral attributes that may not be relevant to our task of acoustic zone localization (e.g., looking at one's shoes, eating with the face down, fiddling with fingers, etc.)~\cite{hendrikse2019movement, frischen2007gaze}. Extracting patterns of interest requires careful construction of the relevant input features and target formulation. Second, IMU sensors suffer from sensor drift, which makes it challenging to rely on longer observation periods for predictions. Leveraging translational information is particularly difficult under these conditions. Finally, the relationship between a speaker talking and a listener orienting themselves in the speaker's direction is not tightly causal, varies across individuals, and is heavily dependent on conversation dynamics~\cite{brimijoin2010auditory, frischen2007gaze}, which makes it generally hard to learn precise acoustic zones purely from head orientation.

\imwut{Given the complex nature of head-orienting behaviors, simple rule-based statistical methods are insufficient to capture natural conversation-focused patterns}. In this work, we show that by constructing suitable proxy tasks, it is possible to infer acoustic zones of interest with minimal prior assumptions. We first localize the acoustic zones of interest under the assumption that the number of conversation partners is known. We then relax this assumption by optimizing a secondary objective that classifies the number of conversation partners. Finally, we integrate these methods into an end-to-end system and demonstrate their robustness under practical conditions, including the availability of abstract audio features—\imwut{coarse speaking-behavior signals such as listener speaking status and partner talkativeness, which are commonly available in speech enhancement systems~\cite{loizou2007speech} and do not require spectrograms or high-dimensional feature processing}—as well as shorter analysis windows and IMU sensor drift. We validate our approach on a large real-world dataset comprising diverse multiparty conversational layouts and natural listening–speaking dynamics. Our key contributions are:

(1) \imwut{We propose a novel task of localizing conversation partners based on listener preferences using head orientation as a behavioral modality, captured from on-device IMUs in smart glasses to support speech-enhancement applications.} We demonstrate the effectiveness of our approach on a large-scale dataset (N > 70) with unconstrained, natural conversations, designed to provide meaningful findings in practical settings.

(2) To address the challenges inherent in behavioral modalities, we propose the \underline{H}ead-orientation-based \underline{A}coustic-zones \underline{Lo}calization (\locnet{}) network, which formulates the prediction of conversation partners’ spatial locations—overlapping with the listener’s acoustic zones of interest—as a sequence-to-multilabel classification problem. \locnet{} achieves an average accuracy of 0.78 and a macro-F1 of 0.62, representing a 24\% average improvement over rule-based and general-purpose baselines. To further reduce reliance on static prior knowledge for localization, we introduce the \underline{C}lassifying the number \underline{o}f \underline{Co}nversation partners (\clfnet{}) network, which attains an average accuracy of 0.74, yielding a 25\% improvement over comparable baselines.

(3) Finally, we present a comprehensive evaluation of our proposed methods and introduce \jointnet{}, an end-to-end training strategy that localizes acoustic zones by estimating the number of conversation partners in a stage-wise manner. We demonstrate the effectiveness of our approach across diverse practical scenarios through extensive ablation studies, qualitative analyses, \imwut{and a minimal speech-enhancement pipeline}, providing intuitive insights into our formulation and adopted methodology.

%% file: sections/relatedworks.tex
\section{Related Works}

\subsection{Linguistic Significance of Head-Orientations}
Psychoacoustic literature extensively explores the communicative and motoric uses of head movement, ranging from its potential to indicate aggressiveness in the speaker~\cite{fitrianie2023head}, hearing and eyesight impairments~\cite{sharkey1990turn}, or speech pathology disorders~\cite{hadar1991speech}, to turn-taking behaviors~\cite{tiede2019noggin,mcclave2000linguistic,duncan1972some}. Past studies~\cite{stiefelhagen2002head} also highlight the phenomenon of undershooting in orienting behavior relative to the speaker's location. Additionally, leveraging this modality presents challenges due to variability in the finite reaction time between orienting behavior and a conversation partner's speaking status, as well as cultural biases~\cite{mcclave2000linguistic}. Most of these studies have been conducted in controlled settings, typically with small datasets (\(N < 10\)) and manual annotations. Drawing inspiration from auditory neuroscience findings, this work aims to develop a data-driven method that leverages head-orienting behaviors to infer a user's auditory attention zones. \imwut{Our multilabel classification formulation, using fixed-length observation segments and discrete acoustic zones as the targets, addresses the delayed responses and undershooting tendencies established in prior behavioral studies~\cite{stiefelhagen2002head,mcclave2000linguistic}.}

\subsection{Methods for Head Orientation Estimation}

Several wearable and IoT applications benefit from continuously monitoring a user's head orientation. Audio-based methods, such as using binaural microphones~\cite{yang2022deepear}, multiple microphone arrays~\cite{felsheim2021head, yang2020soundr}, or even wall reflections~\cite{shen2020voice}, have shown promise in determining the head-orientation of a person. Another class of methods involves using exocentric videos to track and determine head orientation~\cite{hausamann2021evaluation}; however, beyond privacy and energy constraints, visual modalities also suffer from occlusion effects. Some works have explored using egocentric visual data for pose estimation of the wearer~\cite{jiang2021egocentric, tsutsui2018estimating}. In this work, our focus is on leveraging a behavioral modality that conserves power and does not use visual modalities. Some studies have proposed dedicated methods to measure accurate head pose for medical applications, such as cephalometric analyses for diagnosis~\cite{lundstrom1995natural}. Inspired by prior works~\cite{al2021evaluation,bisio2024towards} that have validated head-mounted IMUs as a viable modality for head orientation estimates, we utilize on-device IMU sensors on smartglasses. In this study, we extract approximate head orientation sequences provide an algorithmic solution for identifying a user's acoustic zones of interest with minimal instrumentation overhead.

\subsection{Gaze and Head-orientation-based User Interfaces}

Several studies have explored behavioral modalities, such as gaze~\cite{yi2022gazedock, sidenmark2020bimodalgaze} and head orientation~\cite{sidenmark2019eye, yang2021model}, as user-interface gestures. In particular, head orientation has recently emerged as a promising modality for contextualizing voice-assisted devices~\cite{yang2021model}, functioning as a non-verbal command for earbuds~\cite{apple_airpods, ge2023ehtrack}, human activity recognition~\cite{meyer2022u, stankoski2023recognizing} and overall improving the contextual understanding of intelligent systems~\cite{morency2006head}. Unlike prior work on gesture identification with clear labels, this study seeks to uncover implicit, conversation-driven patterns in natural head movements—which lack direct ground-truth supervision during training—in order to determine a user's acoustic zones of interest, making the task significantly more challenging.


\subsection{Conversation Enhancement using Smartglasses}\label{sec:related_cf}
\begin{wrapfigure}{r}{0.40\linewidth}
    \centering
    \vspace{-10pt} 
    \includegraphics[width=\linewidth]{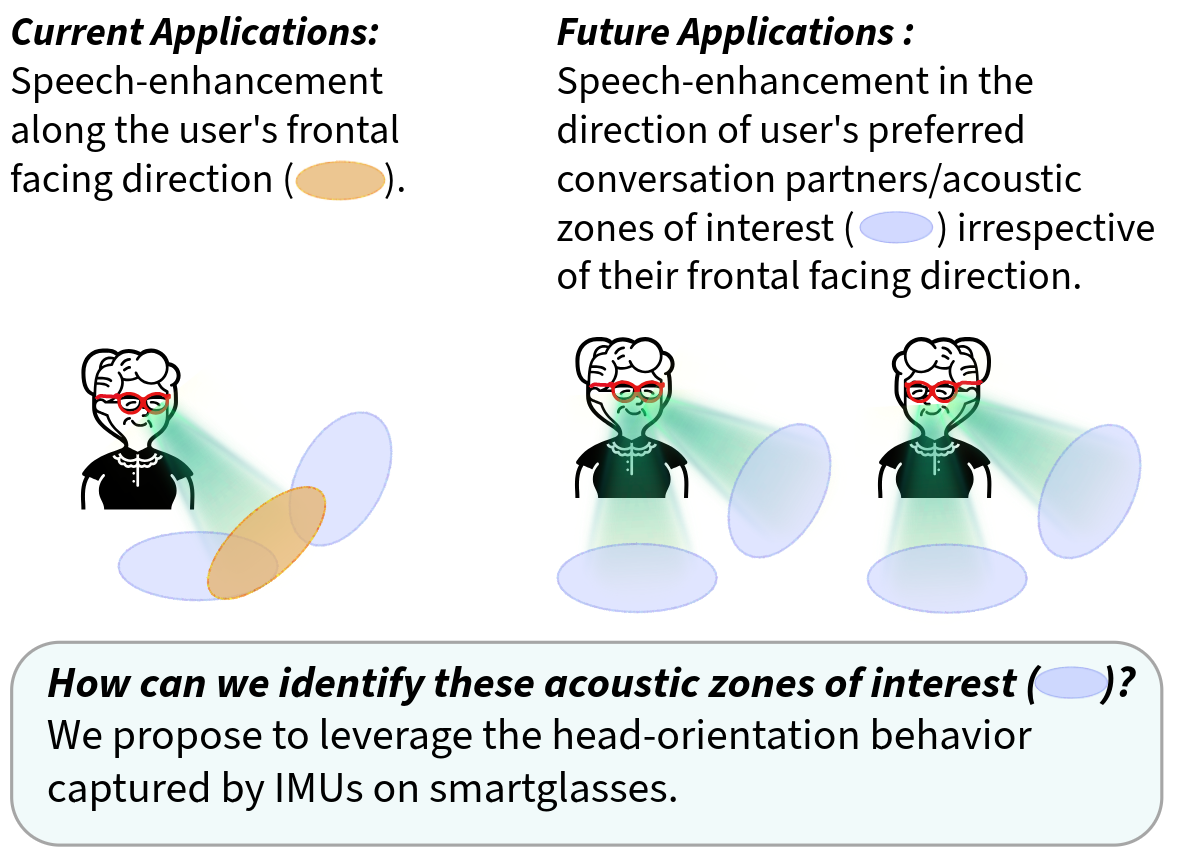}
    \vspace{-20pt}
    \caption{\small Current smartglasses technologies (Left) constrain speech enhancement to the user's frontal-facing direction, disregarding the user's acoustic zones of interest. Our proposed approach (Right) shows that we can use natural head-orienting behavior to identify the acoustic zones of interest, enabling future applications to create a truly immersive conversational experience.} 
    \label{fig:application}
    \vspace{-10pt} 
\end{wrapfigure} An emerging application of smartglasses is to provide conversation enhancement~\cite{zhang2025wearse, xu24i_interspeech, mohapatra2024non}—enhancing sound source from a desired direction and reducing background noise. Currently, this desired direction is primarily determined using two approaches: (1) the user's front-facing direction, or (2) a microphone array on the smart glasses to identify all speech sound sources. The first approach, which uses the user's frontal direction as the conversation-enhancing direction, requires the user to proactively look toward the speaker they wish to focus on, which can be disruptive to the overall conversational experience. The second approach, which relies on a microphone array as described in earlier sections, faces limitations in the presence of interfering speakers and under noisy or non-ideal acoustic conditions. Additionally, for users with hearing difficulties, these existing methods pose substantial challenges, especially when smartglasses are intended to serve as hearing aids themselves~\cite{mehra2020potential} in some applications. This critical limitation presents a compelling opportunity to develop more intuitive methods for inferring the direction in which users want speech enhancement. As illustrated in Figure~\ref{fig:application}, future conversational enhancement technologies must be guided by the user's listening preferences to truly improve the immersiveness of their interaction. In this work, we propose a novel approach using head-orientation-based acoustic zones of interest localization, which addresses the fundamental challenge of identifying a suitable behavioral modality for understanding a user's conversational focus.



%% file: sections/methodology.tex
\section{APPROACH}

In this section, we present our approach, starting with an overview of the study dataset and data preparation steps, followed by an illustrative example to motivate our task formulations and describe our design methodology. We then introduce the acoustic zone localization network, \locnet{}, and the classification network \clfnet{} used to determine the number of conversation partners. For each component, we discuss the associated practical challenges and provide the underlying design rationale.

\subsection{Study Dataset}\label{subsec:motivate}
We used the Reality Labs Research Conversations for Hearing Augmentation Technology (RLR-CHAT) dataset~\cite{yin2024hearing, yun2025spherical, murdock2024self, heo2025gaze} for our study. Researchers collected this data from participants aged 20 to 60 years, including individuals with mild hearing loss. The participants engaged in natural group conversations while seated in arbitrary layouts, with group sizes ranging from 2 to 5 participants. During the conversations, eight loudspeakers surrounded the participants and played cafeteria noise that changed pseudo-randomly every 25–35 seconds at four levels: no noise (quiet), 55, 65, and 75 dBA, covering a range of real-world listening conditions~\cite{wu2018characteristics}. The dataset included a balanced distribution of noise levels.

We used the IMU data streamed from the Aria\footnote{https://www.projectaria.com/} smart glasses~\cite{engel2023project} as our primary input. The IMU is located on the right leg of the glasses and sampled at 1000 Hz. We used an optical motion tracking system, OptiTrack~\cite{OptiTrack, nagymate2018application}, to obtain ground-truth annotations, sampled at 120 Hz. Additionally, to demonstrate the advantages of using simpler audio modalities, we included the speaking and non-speaking states of all participants, which are computationally less expensive to obtain compared to multi-channel audio processing.

To ensure alignment across modalities, we used a manually validated subset of the dataset, similar to previous works~\cite{yun2025spherical, murdock2024self}, which included 71 participants from 36 unique sessions, each lasting approximately 1 hour. Here each session is aligned across different modalities in 30-second segments with standard IMU processing~\cite{engel2023project} to compensate for time-invariant sensor differences across the smart glasses. A fisheye lens recorded video at 5 frames per second, and all modalities are downsampled to this frame rate to further processing to facilitate qualitative validation using a synchronized visual reference. Head rotation frequency of 1 Hz is physiologically considered moderate activity, while rare and vigorous head movements occur at approximately 2.6 Hz~\cite{grossman1988frequency}. Thus, downsampling IMU data to 5 Hz captures the relevant head motion behaviors required for conversation partners location tasks. In this work, we refer to the participant whose vantage point is considered as the \textit{focal user}, while the others are referred to as \textit{conversation partners}. Figure~\ref{fig:dataset_viz} in the Appendix illustrates an overview of our dataset's organization.

Table~\ref{tab:dataset_comparison} presents a comparative summary, highlighting how RLR-CHAT differs from other datasets that have used wearable devices to capture human kinematics in conversational settings. While these datasets vary in size and behavioral tasks, RLR-CHAT stands out as the only one that incorporates IMU streams from wearable smart glasses specifically designed for \textbf{natural conversation scenarios} at a large scale (64 hours). This distinction enables us to investigate deep-learning formulations that leverage head orientation, captured using IMU sensors, as the primary modality for identifying the focal user's acoustic zones of interest.

\begin{table}[htbp!]
\centering
\caption{Summary of conversational datasets utilizing wearable modalities to capture human kinematics across various conversational contexts for \textbf{N} number of participants.}
\resizebox{\textwidth}{!}{%
\footnotesize
\renewcommand{\arraystretch}{1.1}
\setlength{\tabcolsep}{3pt}
\begin{tabular}{@{}p{2.3cm} p{5.5cm} >{\centering\arraybackslash}p{0.8cm} >{\centering\arraybackslash}p{0.8cm} >{\centering\arraybackslash}p{0.9cm} p{2.8cm}@{}}
\toprule
\textbf{Dataset} & \textbf{Form-factor: Kinematics Modality} & \textbf{N} & \imwut{\textbf{Hours}} & \imwut{\textbf{Group Size}} & \textbf{Conversation Context} \\ 
\midrule
MatchnMingle~\cite{cabrera2018matchnmingle} & Badge/Pendant: Tri-axial Acceleration & 92 & \imwut{20} & \imwut{2} & Free-standing speed-dating \\ 
SALSA~\cite{alameda2015salsa} & Badge/Pendant: Tri-axial Acceleration & 18 & \imwut{6} & \imwut{2-3} & Free-standing poster-sessions \\ 
Cattuto et al.~\cite{cattuto2010dynamics} & Badge/Pendant: Radio-Frequency Identification (RFID) & 575 & \imwut{12} & \imwut{2} & One-on-one conversations \\ 
Matic et al.~\cite{matic2012analysis} & Mobile Phone: Tri-axial Acceleration & 50 & \imwut{8} & \imwut{2-4} & Controlled social interactions \\ 
Hung et al.~\cite{hung2013classifying, gedik2018detecting} & Custom Tri-axial Accelerometer & 9 & \imwut{2} & \imwut{2} & Controlled social interactions \\ 
Ferlini et al.~\cite{ferlini2019head} & Earbuds: 6-axis accelerometer and gyroscope & 10 & \imwut{1} & \imwut{2} & Controlled conversation \\
\textbf{RLR-Chat (ours)}~\cite{yin2024hearing, yun2025spherical, murdock2024self} & \textbf{Smart-glasses: 6-axis accelerometer and gyroscope} & \textbf{71} & \imwut{\textbf{64}} & \imwut{\textbf{2-5}} & \textbf{Natural seated conversation} \\ 
\bottomrule
\end{tabular}
}
\label{tab:dataset_comparison}
\end{table}

\subsection{Overview of Data Preparation : Head Orientation Approximation from IMUs} \label{subsec:imu_ho}


The continuous angular velocity, $\boldsymbol{\omega}(t)$, is sampled at discrete regular intervals of $\Delta t$, where the instantaneous angular velocity at a discrete time $t_n = n\Delta t$ is given as, $\boldsymbol{\omega}(t_n) = [\omega_x , \omega_y , \omega_z]^\mathrm{T}$. To estimate the angular displacement of the head, we leverage a simple attitude integration scheme as described by ~\citet{jia2019quaternions} using a quaternion representation \imwut{$\mathbf{q} \in \mathbb{R}^4$}~\cite{sola2017quaternion}. Leveraging key results from~\citet{jia2019quaternions}, we can describe the change in rotation, $\Delta \mathbf{q}$, about the instantaneous axis, $\mathbf{u} = \frac{\boldsymbol{\omega}}{\|\boldsymbol{\omega}\|}$ during $\Delta t$ time units in the IMU's local time frame (rad/seconds in this case) as:

\[
\Delta \mathbf{q} = \cos \frac{\theta}{2} + \mathbf{u} \sin \frac{\theta}{2} = \cos \frac{\|\boldsymbol{\omega}\| \Delta t}{2} + \frac{\boldsymbol{\omega}}{\|\boldsymbol{\omega}\|} \sin \frac{\|\boldsymbol{\omega}\| \Delta t}{2}.
\] 
We can consider the original state as $\mathbf{q}(t)$ rotated to a new state $\mathbf{q}(t + \Delta t) = \Delta \mathbf{q} \, \mathbf{q}(t)$~\cite{jia2019quaternions}. We leverage~\citet{sola2017quaternion}'s results to arrive at the closed form solution for the new rotation, $\mathbf{q}_{t+1}$. A detailed derivation and implementation details, including the overall pseudocode in Algorithm~\ref{alg:rot_update}, are provided in Section~\ref{app:imu_ho_geometry} of the Appendix.

After computing the final rotation matrix (step 26 of Algorithm~\ref{alg:rot_update} in the Appendix), we transform a point, \imwut{$\mathbf{v}_{\text{init\_xyz}}$, in the Cartesian plane to $\mathbf{v}_{\text{fin\_ae}}$} in the spherical coordinate system (as shown in Algorithm~\ref{alg:head_rotation} in the Appendix) using standard techniques~\cite{weisstein2005spherical} under some assumptions: assign $d$, the radius in spherical coordinates, to 1; assume the average front-facing direction of the focal user is the origin, i.e., (azimuth, elevation) = $(0^\circ, 0^\circ)$; and do not model any translational motion. Deriving translational motion from IMUs through standard double integration of tri-axial acceleration data is known to be error-prone~\cite{yan2018ridi}, resulting in significant drift over time. To mitigate this, previous studies have proposed physics-guided, learnable modules for estimating global or relative translational motion~\cite{xu2024mobileposer}. Since our study targets seated conversations and acoustic zones of interest, translational motion offers limited value. Therefore, we focus on leveraging gyroscope measurements over short durations to estimate the head orientation geometry.

\begin{figure}[!htbp]
    \centering
    \includegraphics[width=0.8\linewidth]{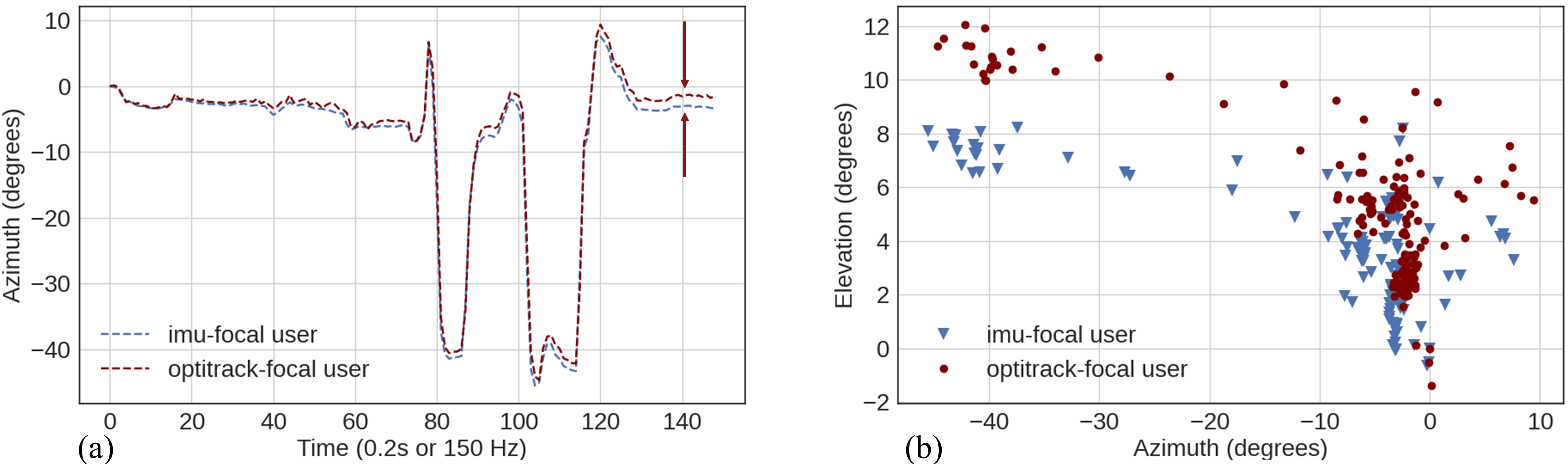}
    \captionsetup{font=small}
    \vspace{-10pt}
    \caption{\small Illustration of a representative case a) showing drift that increases towards the end of the measurement window and b) demonstrates the overall azimuth and elevation from the ground truth optitrack and IMUs.}
    \label{fig:opt_imu_ref}
\end{figure}

Obtaining orientation from gyroscope measurements also incurs drift over time, and correcting for such drift \imwut{often requires device-dependent calibration of sensor-specific bias instability, scale factor errors, and temperature coefficients~\cite{kok2017using,sabatini2011estimating,woodman2007introduction}}. In this work, we explore a device-agnostic way to leverage a modality as a cue for the focal user's auditory zones of interest; therefore, we opt for shorter observation windows in our study. Figure~\ref{fig:opt_imu_ref}(a) shows a comparison for a 30-second segment of the ground-truth head orientation from OptiTrack measurements and the IMU-derived head orientation for the focal user, showing minimal drift toward the end of the segment, supporting our design choice. In a later section (Section~\ref{sec:Opt_IMU_Perf}), our empirical results show that such drift is acceptable in our case. To advance our goal of identifying the focal user's auditory zones of interest, we do not seek a precise head-pose estimate but rather aim to capture the dynamics of the head orientation and its trajectory. Figure~\ref{fig:opt_imu_ref}(b) shows near visual agreement between the IMU- and OptiTrack-derived head orientation for the segment, thus justifying our choice of wearable behavioral modality in this study.



\subsection{Motivation and Rationale} \label{subsec:motivate}
\begin{figure}
    \centering
    \includegraphics[width=\linewidth]{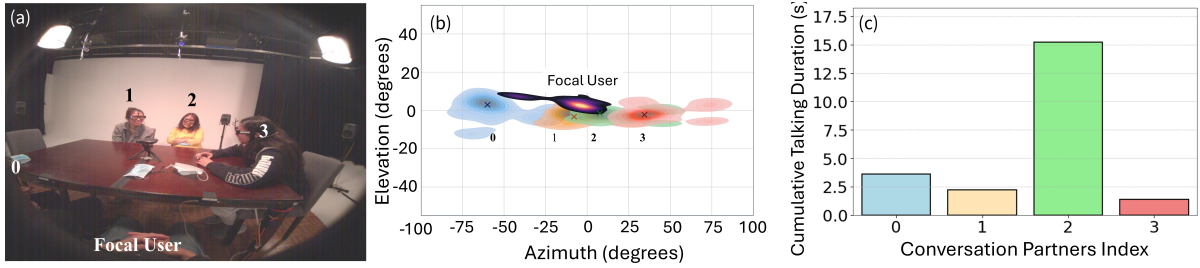}
    \captionsetup{font=small}
    \vspace{-15pt}
    \caption{\small Study Motivation Illustration: (a) Layout of the conversation group, showing the focal user and four conversation partners; (b) Density plot of conversation partners' locations from OptiTrack measurements, transformed to the focal user's frame of reference, with head-orientation data from the IMU on the smartglasses overlaid; (c) Cumulative voice activity of conversation partners during a 30-second segment.}
    \label{fig:motivate}
\end{figure}

Consider a scenario where the focal user is engaged in conversation with four partners, as shown in Figure~\ref{fig:motivate}. We obtain the ground truth locations of the conversation partners from the exocentric OptiTrack~\cite{OptiTrack} cameras that capture the absolute positions of all participants. These fixed cameras track multiple rigid bodies identified by unique markers attached to each participant's head. The rotation matrix derived from OptiTrack measurements is transformed to spherical coordinates (azimuth and elevation) using the standard Algorithm~\ref{alg:head_rotation} provided in the Appendix. Note that the original input from the OptiTrack measurements provides an absolute frame of reference from a third-person point of view. To map to IMU-derived coordinates, which are naturally head-locked or egocentric, similar to past works~\cite{yun2024swl}, we transform the OptiTrack rotation matrix into a head-locked frame of reference. Figure~\ref{fig:motivate}(b) overlays the ground-truth spatial locations of the conversation partners with the focal user’s head orientation, captured by IMU sensors, and shows that the focal user’s acoustic regions of interest align with the locations of the conversation partners. This observation motivates us to leverage head orientation as a behavioral modality to localize conversation partners.

Another observation is that, although there are four conversation partners, the head-kinesis density shows two distinct clusters. Upon further inspection, when we juxtapose the talkativeness of a partner, as shown in Figure~\ref{fig:motivate}(c), we observe a correlation between the talkativeness of a conversation partner and the head-orienting behavior of the focal user—more frequent explicit looking in the direction of conversation partners 0 and 2, who are the more talkative partners in this segment. This motivates us to study methods for understanding another task that can be beneficial in the context of conversation scene understanding: identifying the number of conversation partners based on head orientation behavior.

Overall, this highlights that the head orientation dynamics, even during a seated conversation, are complex and influenced by various contextual factors, not just limited to talkativeness. It also reinforces our choice of the study dataset, which is collected during seated natural conversations, enabling us to be the first to investigate the role of head orientation in localizing and identifying the number of conversation partners with minimal prior knowledge in a systematic manner.

\subsection{Localizing Acoustic Zones of Interest from IMU-derived Head-Orientations} \label{subsec:task1}

In this section, we first outline the approach used to construct the targets, followed by the description of the tasks and the corresponding network design.

\subsubsection{Discrete Spatialization of Acoustic Zones of Interest}\label{subsec:spatialize}

As described in Sections~\ref{subsec:motivate} and~\ref{subsec:imu_ho}, we transform the ground-truth spatio-temporal positions of the conversation partners, obtained from OptiTrack, to the same reference frame as the focal user’s IMU-derived head orientation. Our goal is to identify the acoustic zones of interest, which overlap with the locations of the conversation partners. There are two ways to formulate this: 1) a continuous target prediction, designed as a multi-head regression task. This requires additional considerations for varying group sizes, as the target location of each talker needs to be ordered; or 2) an alternative approach, which discretizes the spatial locations under certain assumptions, framing the problem as a multilabel binary classification task. This latter approach has the advantage of being applicable to any group size without requiring architectural changes. Additionally, the speech-enhancement applications on smartglasses~\cite{xu24i_interspeech, feng2025directional, battula2025robust} that stand to benefit from the integration of behavioral preferences of the focal user also apply their (beamforming) directionality towards discrete spatial zones; supporting our formulation's practical utility. We now present our rationale for choosing this multilabel binary classification formulation.

\noindent \textbf{Rationale for localization task's target construction.} We discretized the spatial locations of the conversation partners—defined as everyone present at the table with the focal user—at the segment level (30 seconds) for the following reasons:\hspace{\fill}\,

\begin{enumerate}
    \item Under nominal conversation settings, a focal user looks in the direction of the intended talker when they speak~\cite{dawson2022your, frischen2007gaze, heo2025gaze}. Our goal is to identify such egocentric behavior patterns from IMUs on smartglasses that are relevant to the conversation and determine the location of acoustic attention. However, the causality between the intended talker's speaking status and head orientation is not deterministic, thus imposing a frame-level prediction (0.2 seconds or 5 Hz) is infeasible for this task. Hence, we determine the locations at a segment level (30 seconds). 
    
    \item From Figure~\ref{fig:motivate} (a), we observe that the intended talkers may undergo some translational motion even during a seated conversation (leaning forward, shifting sideways, etc.). This makes their locations subject to high variability; for instance, the density clusters in Figure~\ref{fig:motivate} (b) are not tightly centered. Considering this, since we cannot couple head orientation with the precise dynamic location of the conversation partners, and estimating the conversation partners' locations so precisely does not offer any benefit in downstream speech-enhancement applications that generally use a fixed beam width of approximately 20$\degree$–60$\degree$~\cite{hladek2019interaction, hoshuyama2002robust, huang2025advances}, we estimate the location statistics over a segment and assign that to the particular talker.  

    \item Sometimes conversation partners are seated very close to each other, and it is known that head orientation exhibits bimodal behavior in one-on-one conversations or undershooting~\cite{hendrikse2019movement, lu2021investigating}, which already makes the task highly challenging. In such scenarios, devising dedicated methods to disambiguate closely seated talkers may not provide significant benefits, following the same reasoning as before: conversation-enhancement applications typically operate with a fixed beam width for signal enhancement~\cite{hladek2019interaction, hoshuyama2002robust, huang2025advances}. Therefore, we propose a multilabel formulation—where each spatial zone may contain more than one conversation partner (described in detail below)—which avoids any rule-based target assignment.

\end{enumerate}

Thus, we propose discrete spatialization of the acoustic zones of interest based on the locations of the conversation partners over a segment, and construct the ground truth as follows.

\noindent \textbf{Constructing Ground-Truth Spatial Locations for Focal User's Acoustic Zones of Interest.} For each speaker \(s\), we compute their median azimuth angle over a segment, \(\theta_s = \text{median}(\Theta^s_t)\), where \(\Theta^s_t\) denotes the set of azimuth angles for speaker \(s\) at each time step \(t\) within the segment, using OptiTrack’s measurements of their true locations. We then generate a bin-vector \(\mathbf{b}_s\) for each speaker \( s \), where each element \( b_{s,i} \) is defined as follows: \begin{wrapfigure}{r}{0.25\textwidth}
    \centering
    \includegraphics[width=\linewidth]{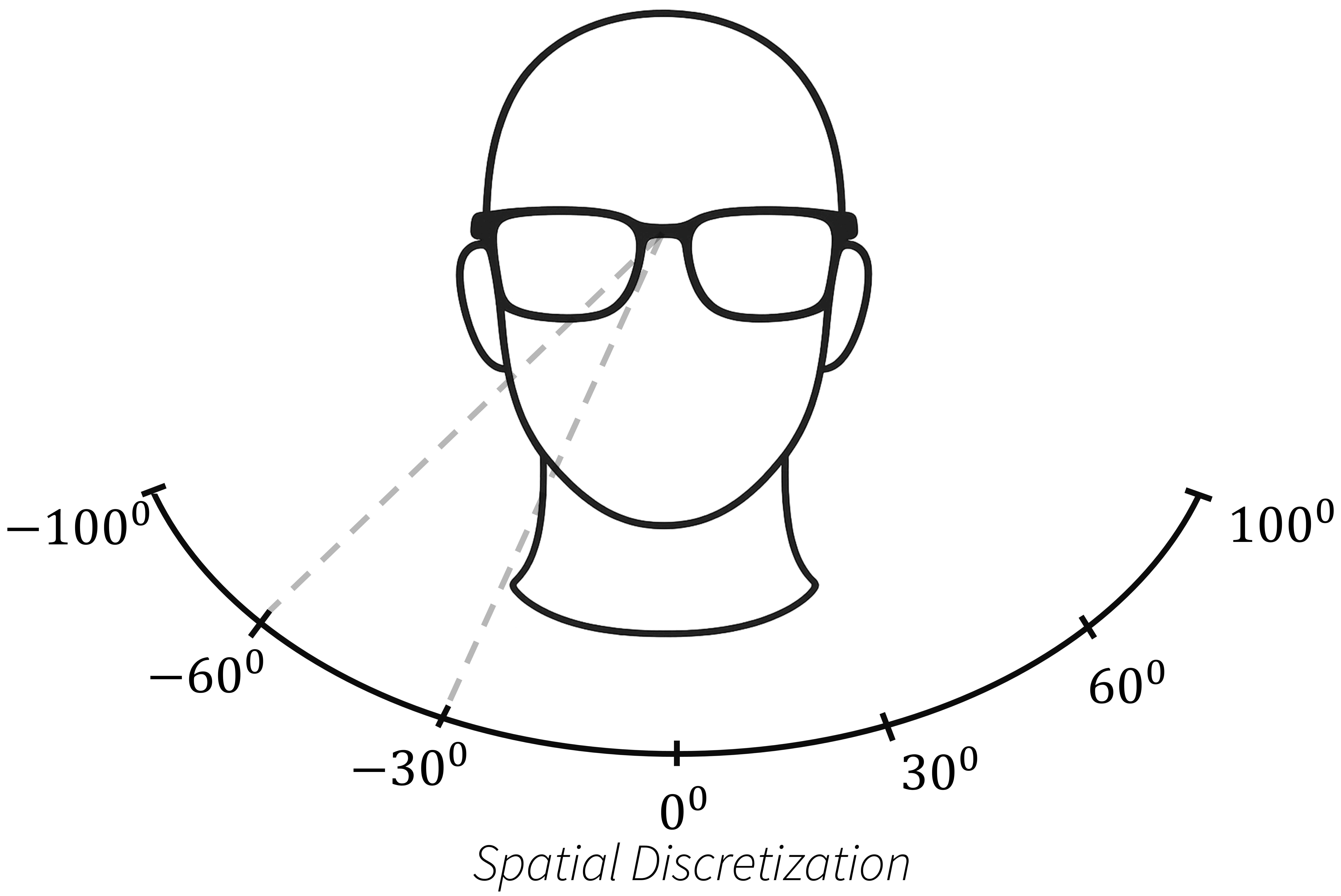}
    \vspace{-15pt}
    \caption{\small Spatial discretization of conversation partner locations. For the layout in Figure~\ref{fig:motivate} (a), the bin vector will be \texttt{[101101]} for these spatial zones.}
    \vspace{-15pt}
    \label{fig:HO_spat_b}
\end{wrapfigure} 
\[
b_{s,i} = 
\begin{cases} 
1, & \text{if } \theta_s \text{ falls within bin } i \\ 
0, & \text{otherwise}
\end{cases}
\]
Here, each bin \( i \in \mathbb{Z}, 0 \leq i \leq 5 \) is defined by the interval \([l_i, r_i]\), where \( l_i \in [100, -60] \) and \( r_i \in [60, -100] \), discretized by an interval \( g_i \), as illustrated in Figure~\ref{fig:HO_spat_b} (\( l_{i+1} = l_i + g_i, \, r_{i+1} = r_i + g_i \)). The bin definitions described by the set \( \tau_{bins} = \{ l_i, r_i, g_i \} \) are selected as hyperparameters (further analysis is given in Section~\ref{sec:Opt_IMU_Perf}) and are not central to our approach. They can be adjusted without loss of generality, depending on the end application. \imwut{The final target bin-vector is computed by performing a logical OR operation across all speakers' bin-vectors: $\mathcal{Z} = \bigvee_{s=1}^{S} \mathbf{b}_s, \quad \mathbf{b}_s \in \{0,1\}^n,$ where $S$ is the number of conversation partners, and $\bigvee$ denotes element-wise logical OR across all $S$ vectors with $n$ bins.}

Such a construction has several advantages: 1) closely seated speakers can be assigned to the same bin, 2) highly animated speakers—those whose movements exceed the bin width across segments—can be assigned to different bins across segments within the same session, and 3) it allows the construction of bin widths in accordance with downstream speech-enhancement applications.


\subsubsection{\locnet{} Network Design for Localizing Acoustic Zones of Interest}\label{subsec:locnet_sec} 
\begin{figure}[!htbp]
    \centering
    \captionsetup{font=small}
    \includegraphics[width=0.75\linewidth]{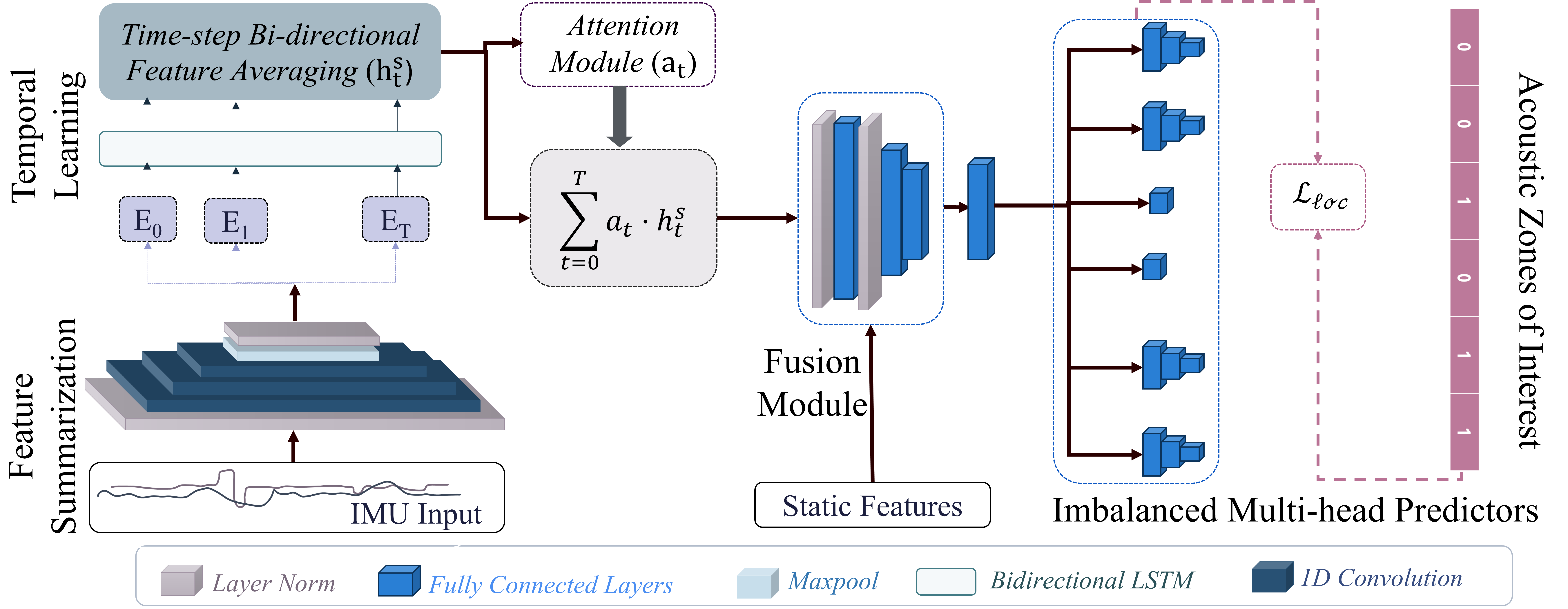}
    \caption{\small Overview of the \locnet{} network used for localizing the acoustic zones of interest: Temporal learning module, followed by the fusion block for incorporating static features, and finally the imbalanced predictors for determining the focal user's acoustic zones of interest.}
    \label{fig:overal_arch}
\end{figure}

Each component of the network, illustrated in Figure~\ref{fig:overal_arch}, is described below. The input is the sequence of spherical coordinates derived in Section~\ref{subsec:imu_ho}, and the target is \(\mathcal{Z}\), as described previously.

\begin{Formulation}{Head-orientation based Acoustic Zones of Interest Localization (\locnet{})}
Given a sequence of head orientation measurements, denoted as \(\mathbf{x}_t\) for all \imwut{\(t \in [0, T)\)}, for a focal user within a group of arbitrary size \(\mathcal{G}\), the objective is to identify the acoustic zones of interest, \(\mathcal{Z} = \{z_1, z_2, \ldots, z_n\}\), where \(n \leq |\mathcal{G}|\). These zones correspond to the spatial locations of one or more conversation partners, inferred from the behavioral data captured by the head orientation sequence. This task is formulated as multilabel classification for a given sequence.
\end{Formulation}
\noindent \textbf{Feature Summarization.} To capture the head-orientation dynamics to solve for the focal user's acoustic zones of interest, we need a sequential learning module. We first normalize the input, $\mathrm{x}_t$ across the features. Then the features are summarized using 1D convolutional neural network layers (CNNs) and applying max pooling along the temporal axis resulting in embeddings, \imwut{ $\mathrm{E} \in \mathbb{R}^{F \times T}$}, where $F$ is the feature-dimension (chosen as a hyperparameter) in the latent space and $T < 150$ for a 30 second segment.

\noindent \textbf{Temporal Learning.} Our task of localizing acoustic zones of interest depends on the ability of the temporal learning module to differentiate between conversation-relevant features and those arising from conversation-agnostic head movements. For example, a focal user might look at their shoes or engage in eating; in such cases, merely analyzing the density of head orientation statistics could incorrectly identify locations like under or on the table as acoustic zones of interest. Our goal is to enable the model to distinguish between behaviors that are not focused on conversation and those that are. Our pilot analysis revealed the ineffectiveness of non-temporal or rule-based methods, and our results in later Sections~\ref{sec:loc_perf} and~\ref{sec:clf_perf} support these design choices with consistent superior performance of \locnet{} over statistical rule-based models and other baselines. We also find transformers to be suboptimal as a sequential backbone in this case as shown in Table~\ref{tab:pilot_seq} (refer to Section~\ref{sec:perf_metrics} for Macro-F1 and Hamming Score definition), possibly due to their limited expressiveness in capturing dependencies across fine-grained time steps similar to prior observations in modeling IMU~\cite{mollyn2023imuposer, xu2024mobileposer, mohapatra2025can} or time-series~\cite{zeng2023transformers, mohapatra2025maestro} data. 
\begin{wraptable}{r}{0.25\textwidth}
\centering
\captionsetup{font=small}  
\caption{A brief analysis of the performance of different sequential backbones for ~\locnet{}.}
\vspace{-5pt}
\label{tab:pilot_seq}
\resizebox{0.25\textwidth}{!}{
  \begin{tabular}{lcc}
  \toprule
  \textbf{Model} & \textbf{Macro-F1} & \textbf{Hamming} \\
  \midrule
  1D CNN        & 0.40 & 0.76 \\
  Transformer   & 0.27 & 0.82 \\
  LSTM          & 0.60 & 0.78 \\
  BiLSTM+Attn   & 0.63 & 0.80 \\
  \bottomrule
  \end{tabular}
}
\vspace{-5pt}
\end{wraptable}

Consequently, we have adopted a variant of the recurrent Long Short-Term Memory (LSTM) for temporal learning in our design. We use a bidirectional LSTM (BiLSTM), where each cell is represented by $\mathcal{L}_\tau$ and parameterized by the set $\tau$. We combine the forward and reverse layers' hidden states, represented as $\overrightarrow{\mathbf{h}}^f$ and $\overleftarrow{\mathbf{h}}^r$, respectively, to obtain $\mathbf{h}^s = \frac{\overrightarrow{\mathbf{h}}^f + \overleftarrow{\mathbf{h}}^r}{2}$, where $\mathbf{h}^s \in \mathbb{R}^{H \times T}$. Then, we apply self-attention similar to \citet{vaswani2017attention}. We initialize the query and key vectors, $Q_a$ and $K_a$, by performing a linear projection of $\mathbf{h}_t$, denoted by $\mathcal{W}$ and parameterized by the indicated subscripts to obtain the self-attention, \imwut{$\mathcal{A} \in \mathbb{R}^{H \times T}$}. The trainable parameters for temporal learning are contained in the set $\tau = \{\xi_h, \xi_q, \xi_k\}$. We then compute the dot product of the attention weights at each time index with the BiLSTM output, taking a weighted mean along the time axis denoted as, $\mathbf{m} = \sum_{t=0}^{T} a_t \cdot \mathbf{h}_t^s$, where $a_t \in \mathcal{A}$, before the fusion block.

\noindent \textbf{Fusion of Static Features.} In addition to the temporal IMU features, we incorporate contextualization of these features with time-invariant information—specifically, the number of total speakers in the conversation setting, which we refer to as static features—through late fusion in \locnet{} to enhance the localization performance. To carry out this fusion in a meaningful manner, we reduce the dimension of the temporally collapsed representation, $\mathrm{m}$, using linear layers $\mathcal{W}(.)$ with the rectified linear unit (ReLU) as the activation function, to obtain $\mathrm{p} \in \mathbb{R}^K$, where $K < H // 4$. This is combined with the static features to produce \(\mathrm{r} \in \mathbb{R}^{K+1}\), which is normalized and passed through linear layers before being input to multi-head predictors. \locnet{}'s has approximately 400k parameters.

\noindent \textbf{Multilabel Classification Objective.} To provide flexibility in weighting the penalty for mispredicting various spatial zones—whether to address a central bias in the frontal direction or to compensate for data imbalance in peripheral locations (refer to Figures~\ref{fig:loc_stat} and ~\ref{fig:spat_loc})—we adopt an imbalanced multi-head binary classifier with deeper classifiers for peripheral zones. Thus, we use a mix of weighted loss functions and imbalanced classifier heads to optimize the task objective. Let the ground truth binary vector for the discretized spatial location being an acoustic zone of interest be denoted as \(\mathbf{b}^{\text{GT}}_i\), and the predicted vector be given as \(\mathbf{b}^{\text{p}}_i = \texttt{ReLU}(\mathcal{W}_{\xi_i}(\mathrm{r}))\) for \(i \in [0, N]\). The weighted cross-entropy loss for the localization task is defined as:

\begin{equation} \label{eq:loc_loss}
    \mathcal{L}_{loc} = -\sum_{i=0}^{N} k_i \left( \mathbf{b}^{\text{GT}}_i \log(\mathbf{b}^{\text{p}}_i) + (1 - \mathbf{b}^{\text{GT}}_i) \log(1 - \mathbf{b}^{\text{p}}_i) \right).
\end{equation}
Here, \(k_i\) is the weight assigned to each acoustic zone based on class imbalance, computed as the inverse of the normalized class frequency, i.e., \(k_i = \frac{1}{\tilde{f}_i}\), where \(\tilde{f}_i = \frac{f_i}{\sum_{j=1}^{N} f_j}\), and \(f_i\) is the number of samples belonging to class \(i\).


\subsection{Number of Conversation-partners from Head-Orientation using IMUs} \label{sec:clf_design}



We are motivated to relax the \textit{a priori} assumption of knowing the number of conversation partners for fusion as static features and explore a methodology to infer it from the head-orientation features. Inspired by the statistical analyses of a focal user's head orientation in a seated conversation setting, our intuition is that, based on the density of the clusters formed by head orientations, as shown in Figure~\ref{fig:motivate} (b), it is reasonable to estimate how many conversation partners are present. We formulate this task as follows,


\begin{Formulation}{Classifying the Number of Conversation Partners (\clfnet{})}
Given a sequence of head orientation measurements and speaking states denoted as \(\mathbf{x}_t\) for all \(t \in [0, L)\), for a focal user, the objective is to identify the number of conversation partners \(\mathcal{G}\). This task is formulated as a sequence-to-one classification problem.
\end{Formulation}

As evident from the illustrative example in Figure~\ref{fig:motivate}, although there are four potential conversation partners, the focal user is mostly oriented towards the central and right-side speakers. This highlights the significant challenge of accurately classifying the number of speakers solely based on head orientation data which is aggravated in larger conversational groups  (discussed in later Section~\ref{sec:Opt_IMU_Perf}'s Figure~\ref{fig:group_acc_clf}). In this section, we describe the design of target shaping and provide details about the additional features and the design of \clfnet{}. Section~\ref{sec:clf_perf} further discusses the challenges and the benefits of incorporating additional modalities (e.g., the focal user's speech state) or domain-specific rules (such as voice-activity-based qualification of conversation partners) for target shaping in the classification performance.

\begin{figure}[!htbp]
    \centering
    \includegraphics[width=0.9\linewidth]{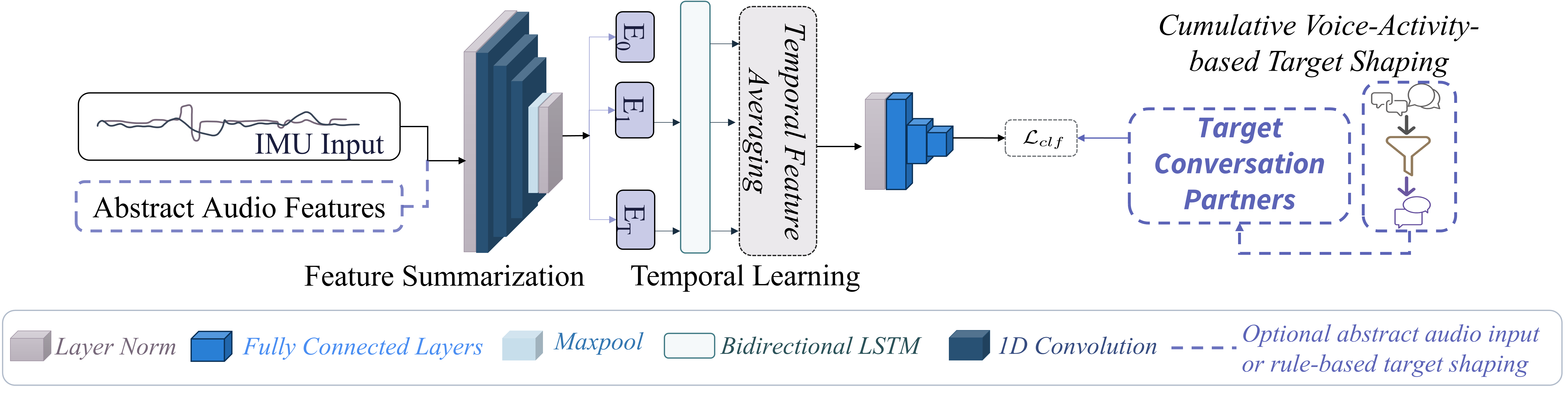}
    \captionsetup{font=small}
    \caption{\small Illustration of the classification network to determine the number of conversation partners based on head-orientation and abstract audio features or target shaping.}
    \label{fig:clf_arch}
\end{figure}

We retain the \textit{Feature Summarization} and \textit{Temporal Learning} modules shown in Figure~\ref{fig:clf_arch}, which serve as backbones for this task. However, in the \textit{Temporal Learning} we did not find self-attention particularly beneficial for this task hence omitted them. The feature embeddings, \(\mathrm{m}\), are now input to a classifier head, \(\mathrm{U}_{clf}\), which predicts the output probability and optimizes the following objective :
$$
\mathcal{L}_{\text{clf}} = -\frac{0}{N} \sum_{i=0}^{N} \sum_{j=1}^{C} y_{ij} \log\left(\frac{\exp(z_{ij})}{\sum_{k=1}^{C} \exp(z_{ik})}\right)
$$

where: \( N \) is the number of samples, \( C \) is the number of classes, \( y_{ij} \) is the binary indicator (0 or 1) if class label \( j \) is the correct classification for sample \( i \), \( z_{ij} \) is the predicted logit for class \( j \) of sample \( i \).

\noindent \textbf{Incorporation of Abstract Audio-features.}
Unlike audio-based speaker localization requiring multichannel audio, here we propose using abstract audio features -- speaking status -- which are cheaper~\cite{price2017low} to obtain. We use two views of speaking status, 1) focal user's voice-activity stream denoted as self\_vad and 2) a logical OR operation on the voice activities of all the speakers to represent an overall conversation partners' voice activity denoted as speaker\_vad. Intuitively, these features help distinguish between speaking-state and listening-state head movements, which can facilitate learning better representations to identify the number of conversational partners.


\noindent \textbf{Cumulative Voice Activity-based Label Shaping.}
Another perspective is to refine the target labels based on the participants' level of talkativeness. Building on our earlier hypotheses—reflecting general human behavior, though not universally applicable—a focal user is more likely to orient their head toward a conversation partner when that partner is actively speaking. However, within a 30-second segment, if certain conversation partners contribute minimally to the conversation, the focal user has limited incentive to look in their direction impacting the performance metrics.

To address this, we adopt a thresholding mechanism, leveraging the voice-activity levels of all conversation partners and accumulating them over a 30-second segment. In the illustrative example in Figure~\ref{fig:motivate}, the talkativeness ranking of the speakers is as follows: Speaker 0 (10\%), Speaker 1 (67\%), Speaker 2 (16\%) and Speaker 3 (6\%). This ranking helps justify the acoustic zones of interest of the focal user. Consequently, we adopt a thresholding mechanism to count the number of participants based on their cumulative talking time within the segment.

\begin{figure}[htbp!]
    \centering
    \begin{minipage}{0.52\textwidth}
        \centering
        \includegraphics[width=0.8\linewidth]{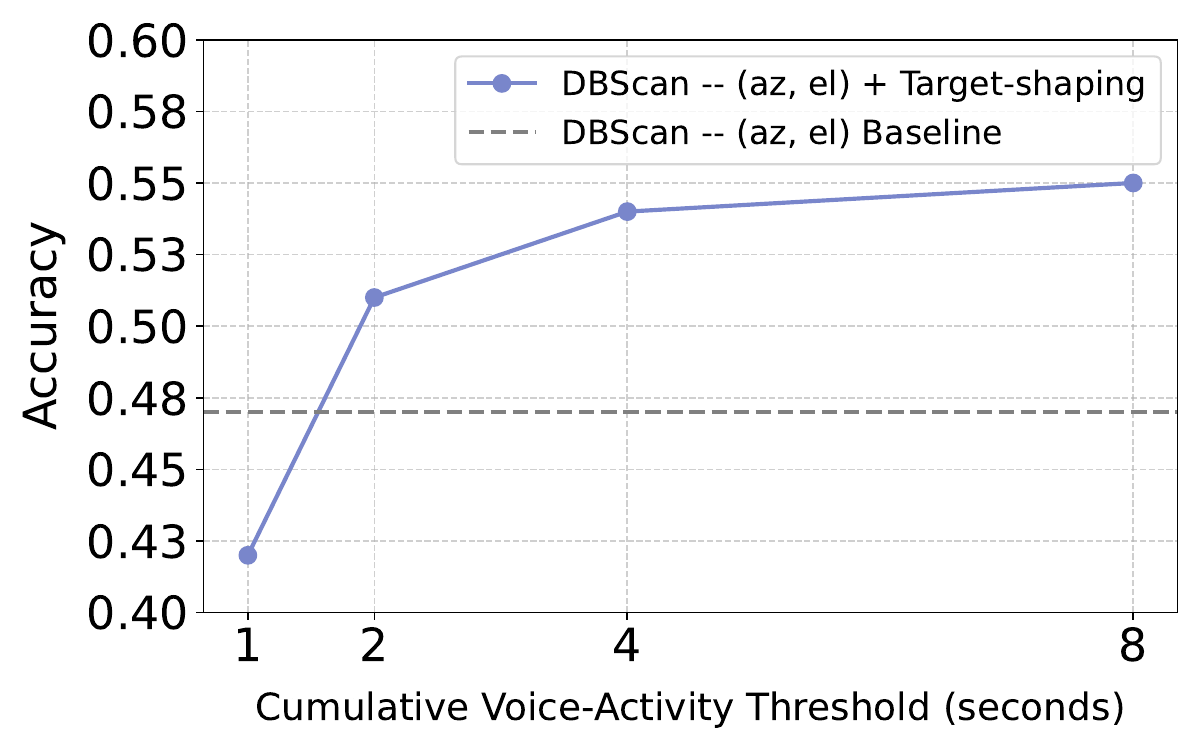}
        \vspace{-10pt}
        \captionsetup{font=small}
        \caption{\small DBSCAN performance in empirically determining a cumulative voice activity threshold to qualify conversation partners and update targets for \clfnet{}.}
        \label{fig:vad_thrshld_acc}
    \end{minipage}%
    \hfill
    \begin{minipage}{0.45\textwidth}
        \centering
        \includegraphics[width=\linewidth]{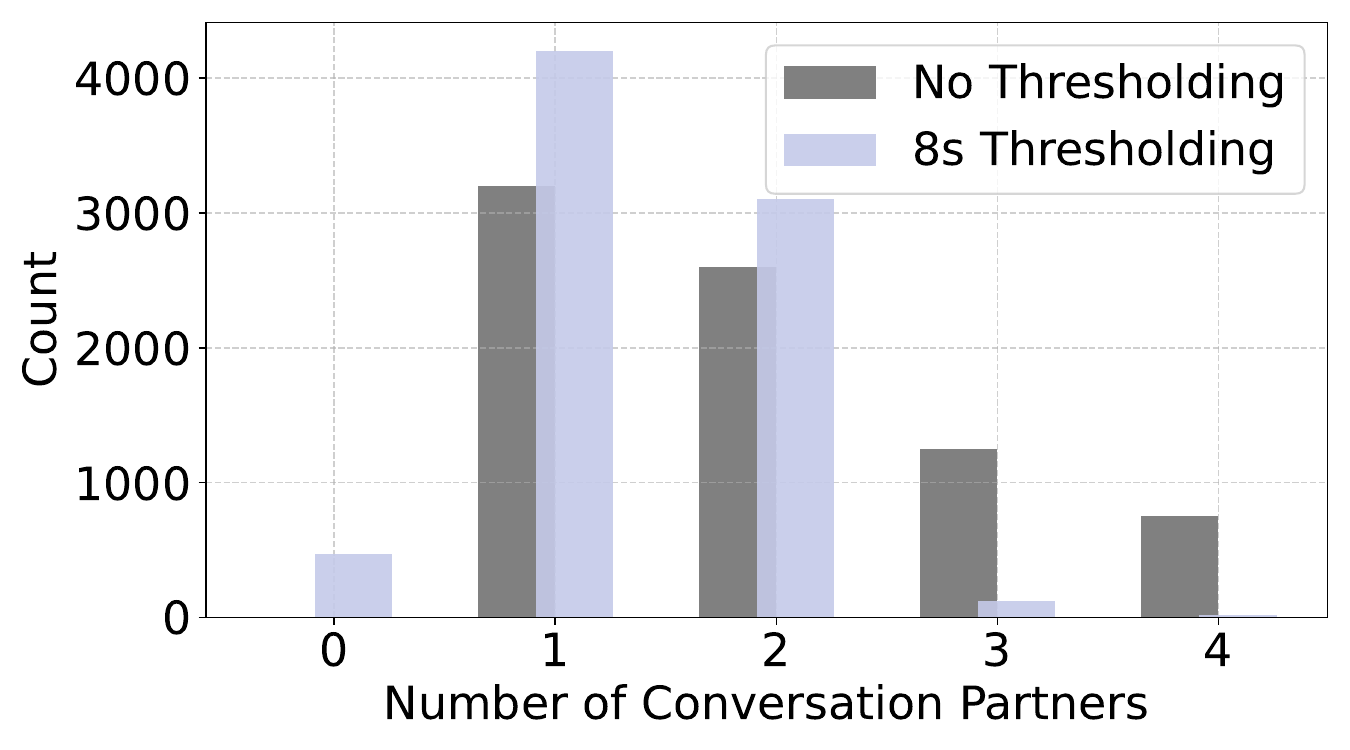}
        \vspace{-20pt}
        \caption{Distribution of the number of conversation partners with and without cumulative voice activity-based label shaping (8s threshold).}
        \label{fig:target_shaping_stat}
    \end{minipage}
\end{figure}


The talkativeness threshold to qualify a participant is determined empirically through a small study using the Density-Based Spatial Clustering of Applications with Noise (DBSCAN) algorithm~\cite{ester1996density}. The distance between points within a cluster and the neighborhood distances that reveal natural spatial groupings are treated as hyperparameters based on the considered threshold. For instance, with a 2-second talkativeness threshold in a 30-second window sampled at 5~Hz, the minimum number of points required to form a valid cluster is set to 10. Based on our pilot analysis, illustrated in Figure~\ref{fig:vad_thrshld_acc}, we selected 8 seconds as the threshold. The original targets and the updated targets after voice-based thresholding are shown in Figure~\ref{fig:target_shaping_stat}. As evident from the target shaping, there are scenarios where none of the speakers qualify as conversation partners, and many segments with more than three targets have been reassigned to a lower target. However, there are still cases where four conversation partners qualify even with the 8-second threshold, as overlapping speakers can occur, and we do not impose any restrictions on such scenarios; we only conduct target shaping based on a speaker's cumulative voice activity.



%% file: sections/evaluation.tex
\section{Results and Discussions}

In this section, we outline our evaluation setup, including the performance metrics and baselines used for comparison. We then present our experimental results demonstrating the performance of \locnet{}, \clfnet{}, and \jointnet{}, which uses estimates from \clfnet{} to reduce the \textit{apriori} for \locnet{} under different conditions. We also present qualitative performance over complete sessions and discuss insights from model explainability. Finally, we present additional experiments to support our design choices.

\subsection{Performance Metrics}\label{sec:perf_metrics}
Following standard recommendations~\cite{read2011classifier, mlkddmulti} for multilabel classification tasks, we adopt the metrics of Hamming Score\footnotetext{\small *It can be viewed as a nuanced form of accuracy in a multilabel formulation.}, Logit-wise Accuracy, Logit-wise F1 and Macro F1 for the localization task. For the task of identifying the number of conversation partners, we leverage the accuracy and Macro F1 metrics following their standard definitions~\cite{read2011classifier}. The formulae for the implemented metrics are provided in Section~\ref{app:perf} of the Appendix.

\begin{figure}[!htbp]
    \centering
    \begin{minipage}{0.48\textwidth} 
        \centering
        \includegraphics[width=0.8\linewidth]{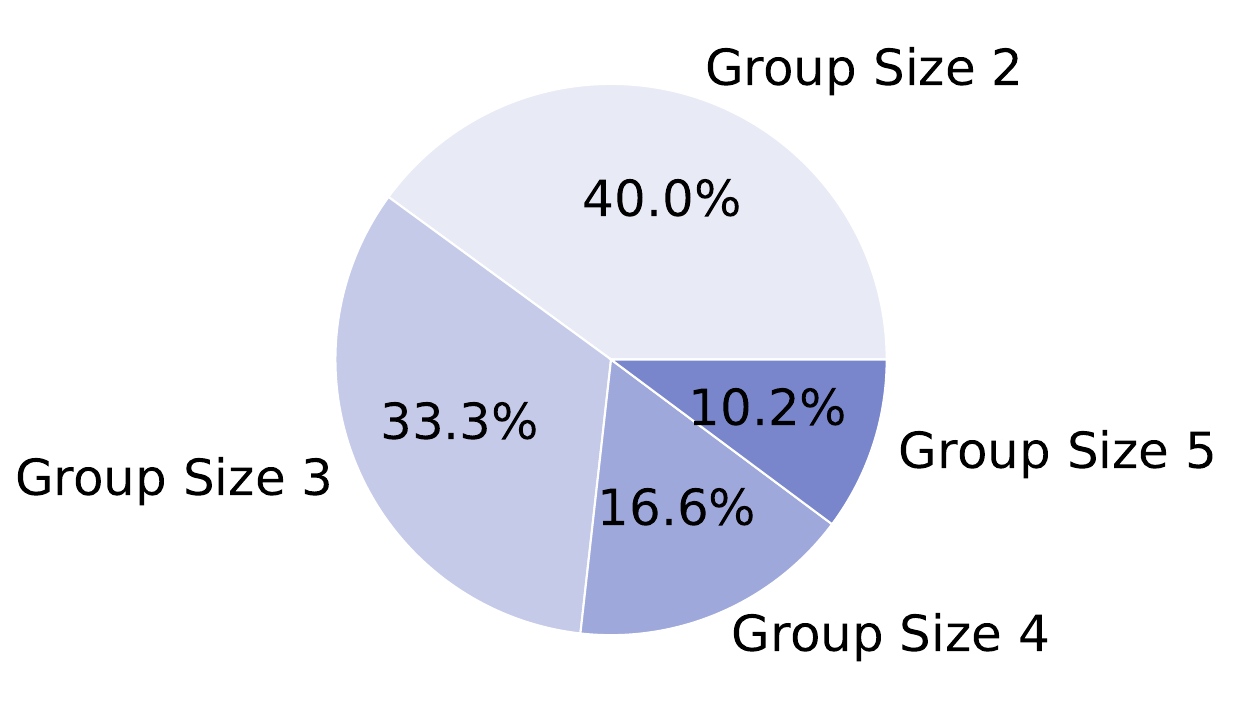}
        \captionsetup{font=small}
        \caption{\small Distribution of group sizes across all sessions.}
        \label{fig:clf_stat}
    \end{minipage}%
    \hfill
    \begin{minipage}{0.48\textwidth} 
        \centering
        \includegraphics[width=0.8\linewidth]{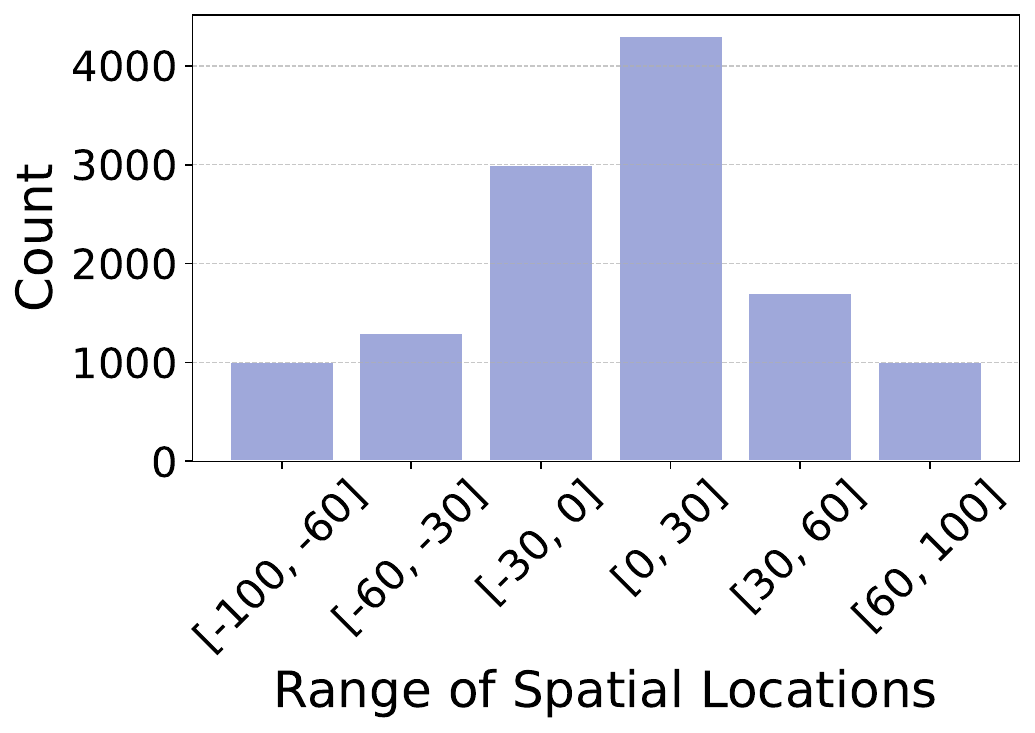}
        \vspace{-10pt}
        \captionsetup{font=small}
        \caption{\small Spatial distribution of conversation partners from the focal user's perspective across all sessions.}
        \label{fig:loc_stat}
    \end{minipage}
\end{figure}

\noindent \textbf{Data Statistics.} The statistics for class distribution and a representative discrete spatialization (refer to Section~\ref{subsec:spatialize}) performed in this study are shown in Figures~\ref{fig:clf_stat} and~\ref{fig:loc_stat}, respectively. 

\subsection{Baselines}\label{sec:baselines}
We carry out evaluations for both the tasks—localization and identifying the number of conversation partners—against three baseline methods \imwut{on the RLR-Chat dataset}. First, we consider a rule-based method, engineered based on knowledge of patterns in conversation dynamics. Next, we use a simple multi-layer perceptron-based deep learning model, which does not account for temporal ordering. Finally, we employ a state-of-the-art time-series representation learning method, Informer~\cite{zhou2021informer}, designed for handling long sequences which takes the downsampled 6-axis IMU data.

\begin{wrapfigure}{r}{0.4\linewidth}
    \centering
    \vspace{-10pt} 
    \includegraphics[width=\linewidth]{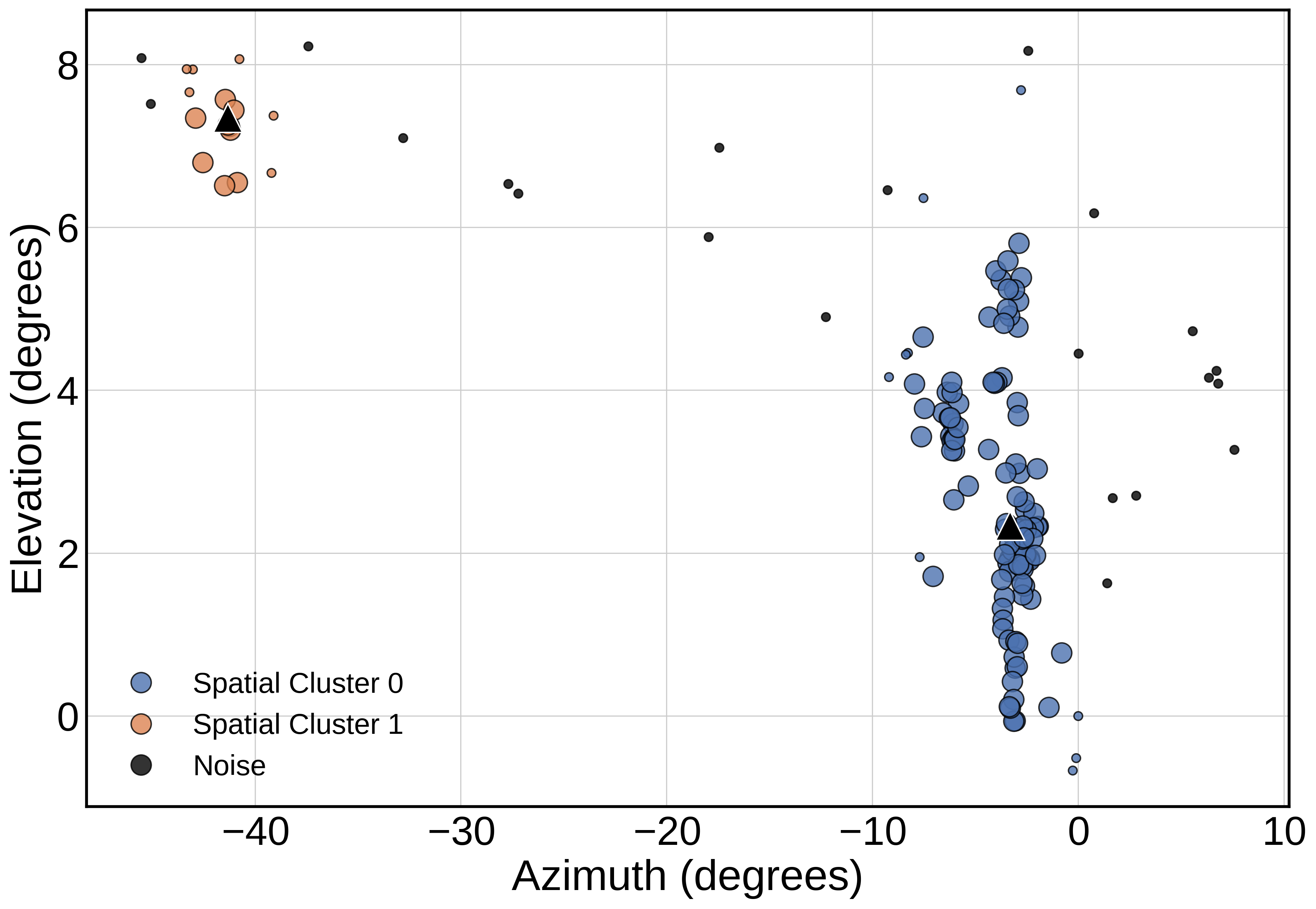}
    \vspace{-10pt}
    \caption{Performance of the rule-based method: spatial density-based clustering of the focal user's head orientation during their non-speaking state within a segment, demonstrated for a representative setting (same scene as shown in Figure~\ref{fig:motivate}). The method identifies two conversation partners and localizes them, following the discrete spatialization strategy in Section~\ref{subsec:spatialize}, to the $[-30\degree, -60\degree]$ and $[0\degree, -30\degree]$ zones in the azimuthal plane.}
    \label{fig:dbscan_demo}
    \vspace{-10pt} 
\end{wrapfigure}

\noindent \textbf{Rule-based.} This is a simplified rule-based method to capture head-orientation dynamics, inspired by previous works~\cite{tan2021multimodal, rienks2005differences} that model the visual focus of attention based on speaking status, head orientation, and other contextual information~\cite{ba2008recognizing}. Other works~\cite{mcclave2000linguistic} have shown, through manual statistical analyses in dyadic conversations, that there is a high correlation of direction of facing with the conversation partner's location, especially during the listening state. Although these tasks are not exactly what we are trying to accomplish, inspired by these findings, we can design a rule-based non-parametric scheme to capture a plausible dynamic: \imwut{when the focal user is not speaking, they look at the actively speaking conversation partner}. Thus, for each segment during the non-speaking state of the focal user, based on the density of the head-orientation coordinates, we conduct spatial clustering (following the DBSCAN algorithm). For the localization task, we take the centroids of the clusters as the spatial zones of interest, and for identifying the number of conversation partners, we count the number of naturally formed clusters. An illustration of this baseline's working principle, applied to the layout previously demonstrated in Section~\ref{subsec:motivate} and Figure~\ref{fig:motivate}, is shown in Figure~\ref{fig:dbscan_demo}.

\noindent \textbf{Segment-based.} This is segment-level non-temporal approach (of approximately 200k parameters) using the azimuth information from the head-orientation computation. We feed the \(L\)-dimensional azimuth as input vector to a multi-layer perceptron (MLP) with: 1) \(n\) classifier heads for the multilabel classification of the localization of acoustic zones of interest task, and 2) one classifier head for the task of identifying the number of conversation partners.

\noindent \textbf{Temporal-based.} Informer~\cite{zhou2021informer} is a transformer-based architecture specifically designed to handle long sequences encountered in the real world, enabling the model to manage long-range dependencies with computational efficiency by using sparse-probability attention. We leverage the encoder part of this architecture and take the mean along the temporal axis before feeding these learned embeddings to task-specific classifier heads~\cite{mohapatra2025can}. The input to this baseline is the downsampled 100 Hz 6-axis IMU data and the model has of approximately 900k parameters. We adopt this baseline to represent a time-series model with high temporal modeling capacity.

\noindent \textbf{Implementation Details.} All experiments are conducted using three random seeds (2711, 2712, 2713), reporting the mean and standard deviation of metrics. Data is split into training and testing sets in a 7:3 ratio, with 20\% of the training data reserved for validation. Models are trained using a batch size of 64 and the ADAM optimizer. Localization models use a learning rate of $1\mathrm{e}{-5}$, while classification models use $1\mathrm{e}{-3}$. All models are trained for 20 epochs, and the best checkpoint is selected based on the lowest validation loss.

\subsection{Evaluating the Localization of Acoustic Zones of Interest}\label{sec:loc_perf}

In this section, we demonstrate the performance of \locnet{} in localizing a focal user's acoustic zones of interest, analyze the influence of various input features, and present qualitative analyses on longer session-wise inputs and the explainability of \locnet{}'s predictions.

\begin{table}[!htbp]
    \centering
    \caption{Performance of the methods for localizing auditory zones of interest, evaluated using logit-wise F1, logit-wise accuracy, macro F1, and Hamming scores (defined in Section~\ref{app:perf} of the Appendix). The best performance is shown in \textbf{bold}, and the second-best is \underline{underlined}.
    }
    \vspace{-10pt}
    \label{tab:loc_f1}
    \resizebox{0.9\textwidth}{!}{
    \begin{tabular}{l|cccccc|c}
        \toprule
        &
        \multicolumn{6}{c|}{\textbf{Discrete-Spatial Zones}} &
        \textbf{Aggregate} \\
        \cmidrule(lr){2-7}
         & \textbf{[-100, -60]} & \textbf{[-60, -30]} & \textbf{[-30, 0]} & \textbf{[0, 30]} & \textbf{[30, 60]} & \textbf{[60, 100]} & \textbf{Scores} \\
        
        \cmidrule(lr){2-8}
        \textbf{Method} & \multicolumn{6}{c|}{\textbf{Logit-wise F1}} & \textbf{Macro F1} \\
        \midrule

        Rule-based &
        0.02 & 0.26 & 0.57 & 0.62 & 0.15 & 0.03 & 0.27 \\
        
        Segment-based &
        0.03\textsubscript{\(\pm\)0.05} & 0.08\textsubscript{\(\pm\)0.08} & 0.58\textsubscript{\(\pm\)0.05} &
        \underline{0.76}\textsubscript{\(\pm\)0.05} & 0.10\textsubscript{\(\pm\)0.01} & 0.07\textsubscript{\(\pm\)0.02} & 0.27\textsubscript{\(\pm\)0.01} \\
        
        Informer &
        \underline{0.31}\textsubscript{\(\pm\)0.01} & \underline{0.21}\textsubscript{\(\pm\)0.03} & \underline{0.62}\textsubscript{\(\pm\)0.08} &
        0.60\textsubscript{\(\pm\)0.12} & \textbf{0.51}\textsubscript{\(\pm\)0.07} & \underline{0.48}\textsubscript{\(\pm\)0.04} & \underline{0.45}\textsubscript{\(\pm\)0.04} \\

        \rowcolor[HTML]{E0E3F5}
        \locnet{} (Ours) &
        \textbf{0.57}\textsubscript{\(\pm\)0.04} & \textbf{0.57}\textsubscript{\(\pm\)0.04} & \textbf{0.65}\textsubscript{\(\pm\)0.01} &
        \textbf{0.80}\textsubscript{\(\pm\)0.01} & \underline{0.42}\textsubscript{\(\pm\)0.03} & \textbf{0.61}\textsubscript{\(\pm\)0.11} & \textbf{0.62}\textsubscript{\(\pm\)0.03} \\
        
        \midrule
        \textbf{Method} & \multicolumn{6}{c|}{\textbf{Logit-wise Accuracy}} & \textbf{Hamming Score} \\
        \midrule

        Rule-based &
        \underline{0.84} & 0.81 & 0.50 & 0.53 & \underline{0.75} & \underline{0.82} & 0.71 \\
        
        Segment-based &
        0.56\textsubscript{\(\pm\)0.48} & 0.59\textsubscript{\(\pm\)0.38} & 0.52\textsubscript{\(\pm\)0.03} &
        \textbf{0.69}\textsubscript{\(\pm\)0.10} & 0.53\textsubscript{\(\pm\)0.37} & 0.76\textsubscript{\(\pm\)0.08} & \underline{0.73}\textsubscript{\(\pm\)0.02} \\

        Informer &
        0.82\textsubscript{\(\pm\)0.01} & \underline{0.83}\textsubscript{\(\pm\)0.03} & \underline{0.62}\textsubscript{\(\pm\)0.00} &
        0.54\textsubscript{\(\pm\)0.06} & 0.64\textsubscript{\(\pm\)0.06} & 0.63\textsubscript{\(\pm\)0.13} & 0.68\textsubscript{\(\pm\)0.02} \\

        \rowcolor[HTML]{E0E3F5}
        \locnet{} (Ours) &
        \textbf{0.88}\textsubscript{\(\pm\)0.01} & \textbf{0.85}\textsubscript{\(\pm\)0.03} & \textbf{0.65}\textsubscript{\(\pm\)0.02} &
        \underline{0.67}\textsubscript{\(\pm\)0.00} & \textbf{0.81}\textsubscript{\(\pm\)0.02} & \textbf{0.86}\textsubscript{\(\pm\)0.02} & \textbf{0.79}\textsubscript{\(\pm\)0.01} \\
        \bottomrule
    \end{tabular}
    }
\end{table}

Table~\ref{tab:loc_f1} summarizes the performance of the rule-based, segment-based, and representative high-capacity time-series baselines against \locnet{}. We observe that: 1) the rule-based model, despite incorporating the focal user's speaking status as an additional attribute, performs poorly in the localization task (0.27 macro-F1 and a Hamming score of 0.71). This aligns with prior studies that relied on heuristics and have noted the inherent complexity of using behavioral modalities to infer auditory attention~\cite{lu2021investigating}; 2) non-temporal models, both rule-based and deep learning-based, are suboptimal in capturing the complex head-orientation dynamics that are indicative of the focal user's attention. This supports our sequential modeling design choice of \locnet{}, which boosts average performance from 0.27 to 0.62 in terms of macro-F1, and from 0.72 to 0.79 in terms of Hamming score over non-temporal models; and 3) direct application of IMU data to state-of-the-art high-capacity sequential models like Informer is not suitable. Our specialized \locnet{}, which incorporates sequential fusion of static features, outperforms Informer by an average relative improvement of 38\% in F1 score and 17\% in Hamming score. Overall, our proposed sequential fusion-based \locnet{} outperforms all baseline methods by an average of 45\% in macro-F1 and 10\% in Hamming score. Intuitively, our approach of off-loading some of the filtering of useful semantics by preprocessing to obtain the azimuth and elevation, followed by stage-wise feature summarization, sequential learning, and a fusion module, supports achieving the high performance in the localization of discrete-spatial zones using behavioral attributes.

\begin{wrapfigure}{r}{0.5\textwidth}
    \centering
    \includegraphics[width=0.95\linewidth]{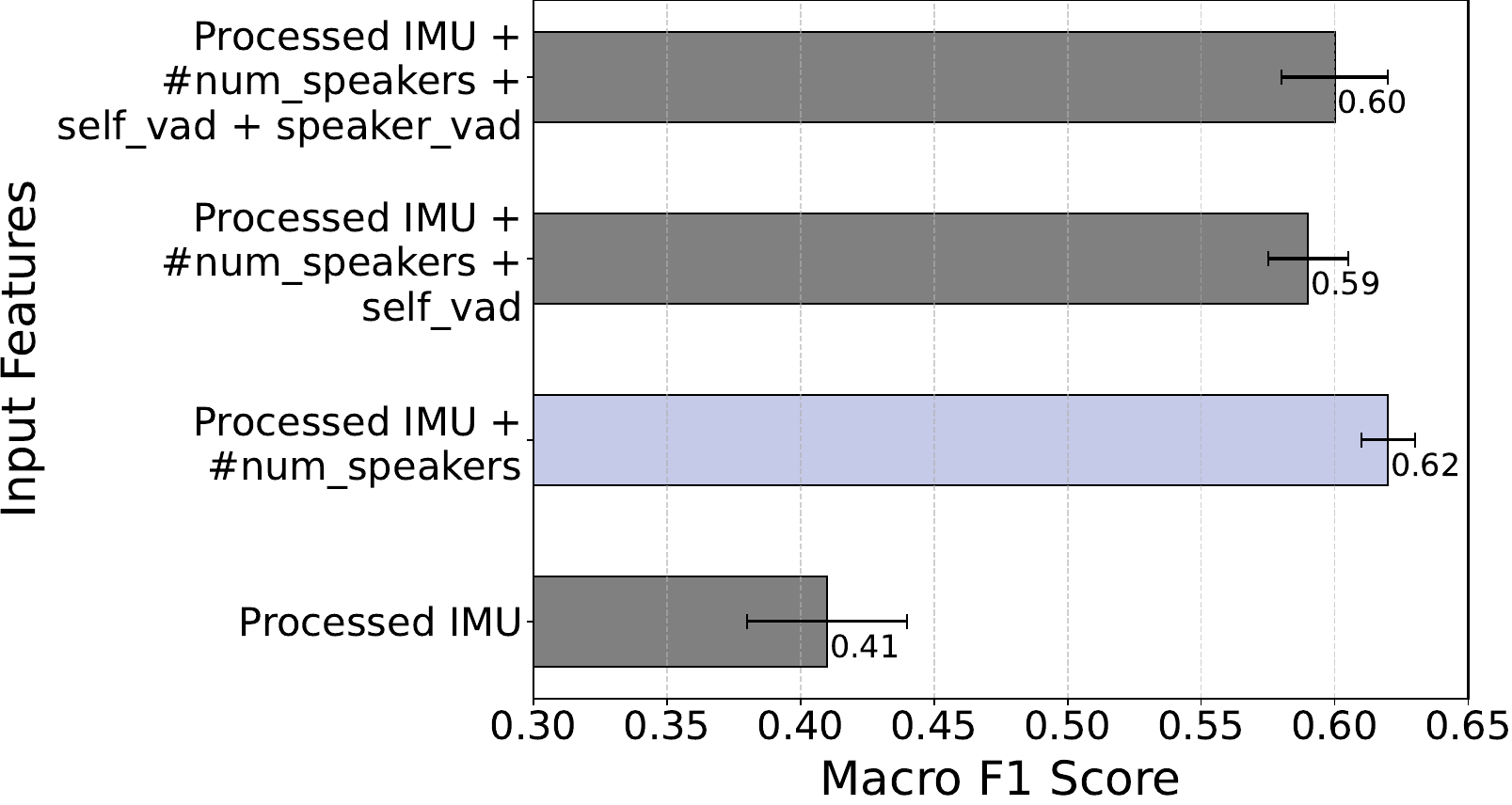}
    \vspace{-10pt}
    \caption{Performance of conversation partners' localization task with different input features. Knowing the number of speakers \textit{a priori} provides a significant performance boost compared to adding other audio-modality-derived features, such as the self or speakers' speech status.}
    \vspace{-10pt}
    \label{fig:loc_importance}
\end{wrapfigure}\noindent \textbf{\textit{A priori} knowledge of the number of speakers enhances localization performance.} Figure~\ref{fig:loc_importance} illustrates the impact of different input features along with the processed IMU features on localization performance. We observe that the late fusion of the number of conversation partners enhances the localization performance by approximately 50\%, as it helps in contextualizing the predictions better. In later Sections~\ref{sec:clf_perf} and~\ref{sec:clf_loc_comb}, we evaluate strategies to obtain this static feature of the number of conversation partners as an embedding from the~\clfnet{} network. We also evaluate the value of incorporating more affordable and abstract audio modalities, such as a stream of the focal user's speaking status denoted as self-voice-activity-detected (self\_vad) and a binarized stream of voice activity from any of the speakers, known as speaker-voice-activity-detected (speaker\_vad). We find that these additional features do not provide any significant benefit for this task.


\begin{figure}[!htbp]
    \centering
    \includegraphics[width=0.7\linewidth]{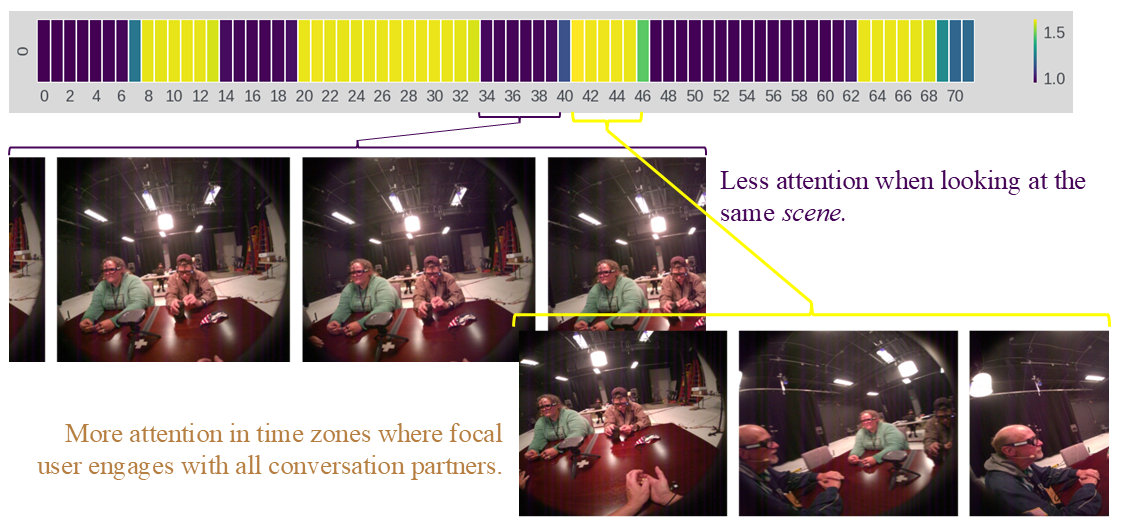}
    \captionsetup{font=small}
    \caption{\small Illustration of the model's temporal self-attention mechanism for predicting conversation partners' locations based on head orientation from IMU data. The mechanism assigns higher weights when the focal user's head orientation exhibits greater dynamism. Visual data is used solely for interpretability analysis.}
    \label{fig:attn_viz}
\end{figure}
\noindent \textbf{Model explainability indicates more attention during active engagement of focal user.} To interpret our proposed predictive model, we extract the attention vector from the \textit{Attention Module} of the model (Figure~\ref{fig:overal_arch}). We normalize the attention vector $a_t \in \mathbb{R}^{H \times T}$ and approximately align the video data to its length to facilitate interpretability. Due to network non-linearities, a precise time-scale mapping is infeasible, but this approximate alignment~\cite{pandey2026timesliver} facilitates the visualization of the temporal importance attributed to each frame in the vision data for intuitive understanding. Figure~\ref{fig:attn_viz} illustrates a representative example, which shows higher attention values when there is more \textit{dynamism} in the head orientation, i.e., the focal user looks in the direction of all the conversation partners, and less when they stagnate their attention. This observation aligns with the expectation that the trajectory of head orientation can help in inferring the conversation partners' locations and further reinforces the value of preserving the temporal dynamics of the head orientation for auditory attention localization task.

\noindent \textbf{Qualitative results across full-sessions show promising localization performance.} We evaluate the overall performance of the identified conversation zones for a complete session. The RLR-Chat dataset consists of sessions which are manually aligned and split into non-consecutive but ordered 30-second segments (refer Figure~\ref{fig:dataset_viz}. (b) in the Appendix).  We note that although the data is collected in a seated-conversation setting, one might assume that the ground truth locations of the conversation partners, after discretization in the azimuthal plane, remain fixed across all segments for a given egocentric session, that is not the case. Due to translational movements during more animated conversations (such as laughing or lateral oscillation) may result in shifted ground truth median locations. This shift affects the ground truth bin vector, $\mathbf{b}^{\text{GT}}_i$, for the $i$-th zone as:
\[
\mathbf{b}^{\text{GT}}_i =
\begin{cases}
[0, \mathbf{b}^{\text{GT}}_0, \mathbf{b}^{\text{GT}}_1, \ldots, \mathbf{b}^{\text{GT}}_{N-1}] & \text{(right shift)} \\
[\mathbf{b}^{\text{GT}}_1, \mathbf{b}^{\text{GT}}_2, \ldots, \mathbf{b}^{\text{GT}}_N, 0] & \text{(left shift)}.
\end{cases}
\]

\begin{figure}
    \centering
    \includegraphics[width=0.85\linewidth]{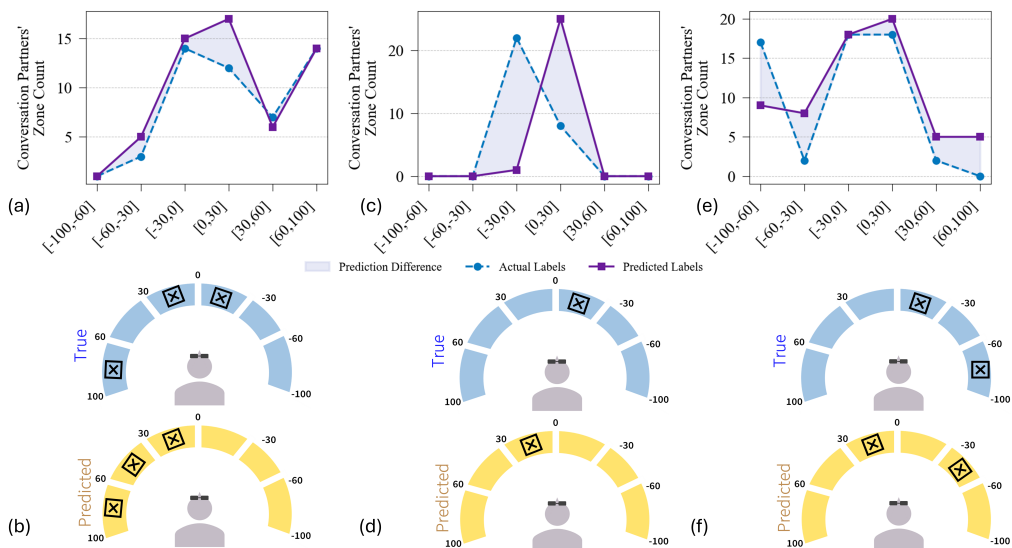}
    \captionsetup{font=small}
    \vspace{-10pt}
    \caption{\small (a, b) Sandwiched Mispredictions; (c, d) Shifted prediction for a single conversation partner; (e, f) Shifted prediction for a two conversation partners.}
    \vspace{-15pt}
    \label{fig:qual_loc}
\end{figure}


For each session, we obtain a count for each spatial zone being a true location of a conversation partner, given by $\mathrm{c}^{\text{GT}}_i = \sum_{m=1}^{M} \mathbf{b}^{\text{GT}}_i(m)$, where $\mathbf{b}^{\text{GT}}_i(m)$ is the $m$-th element of the ground-truth bin vector for the $i$-th segment and $M$ is the number of elements in the bin vector. Similarly, we obtain the session-wise count vector $\mathrm{c}^{\text{P}}_i$ from the predicted bin vectors for each segment $\mathbf{b}^{\text{P}}_i(m)$.

Quantitatively, an exact match between these count-vectors,$\mathrm{c}^{\text{GT}}_i$ and $\mathrm{c}^{\text{p}}_i$ is unsuitable because head orientation is a behavioral modality and does not contain immediate causality of the speaker's motion and focal-user's head movement. Despite challenges such as the influence of conversational context on head orientation, our investigation demonstrates that aggregating predictions from 30-second segments over the course of one-hour sessions yields overall positive results. As illustrated in Figure~\ref{fig:qual_loc}, while certain segment-level mispredictions occur, the model’s session-level predictions remain largely consistent with the ground-truth acoustic zones of interest. We highlight two representative cases: (1) mispredicting a conversation partner positioned between two others, and (2) slightly undershooting the true location of the conversation partners.

Figure~\ref{fig:qual_loc}. (a) illustrates the first case in a four-person conversation setting, where the session-wise prediction's mean absolute error is 1.5, with a high Pearson correlation~\cite{Pearson1895} of 0.94. This indicates that head orientation behavior can reliably inform about the acoustic zones of interest of a focal user over the duration of a full-length conversation, with only short-segment level modeling to accommodate unobtrusive sensing by IMUs on smartglasses. However, at a segment level, like ~\ref{fig:qual_loc}. (b), it may mispredict a \textit{sandwiched} zone between the location zones of two ground truth conversation partners. This can be attributed to the segment-to-segment variability in the conversation partners' median location and overall behavior-driven variability of the IMU-derived head orientation data, which embeds a non-deterministic causality between the head orientation of the focal user and the momentarily updated location of the conversation partner. It also highlights the overall complexity of this task.

In Figures~\ref{fig:qual_loc}. (c), and~\ref{fig:qual_loc}. (e), we illustrate a one-on-one conversation and a three-party conversation setting respectively where undershooting behavior of the head movement as noted by past works~\cite{lertpoompunya2024head, murdock2024self} is evident. Representative segments from these sessions in Figures~\ref{fig:qual_loc}. (d, f) indicate this left-biased undershooting behavior of the focal user.

\subsection{Performance of Identifying the Number of Conversation Partners} \label{sec:clf_perf}

Figure~\ref{fig:loc_importance} highlights that having an \textit{a priori} estimate of the number of speakers in a conversation significantly enhances localization performance. In this section, we evaluate the performance of the proposed \clfnet{} in identifying the number of conversation partners under different settings, compared to other baselines.


\begin{figure}[!htbp]
    \centering
    \resizebox{0.9\textwidth}{!}{ 
    \begin{minipage}{\textwidth}
        \begin{subfigure}[b]{0.48\textwidth} 
            \centering
            \includegraphics[width=\linewidth]{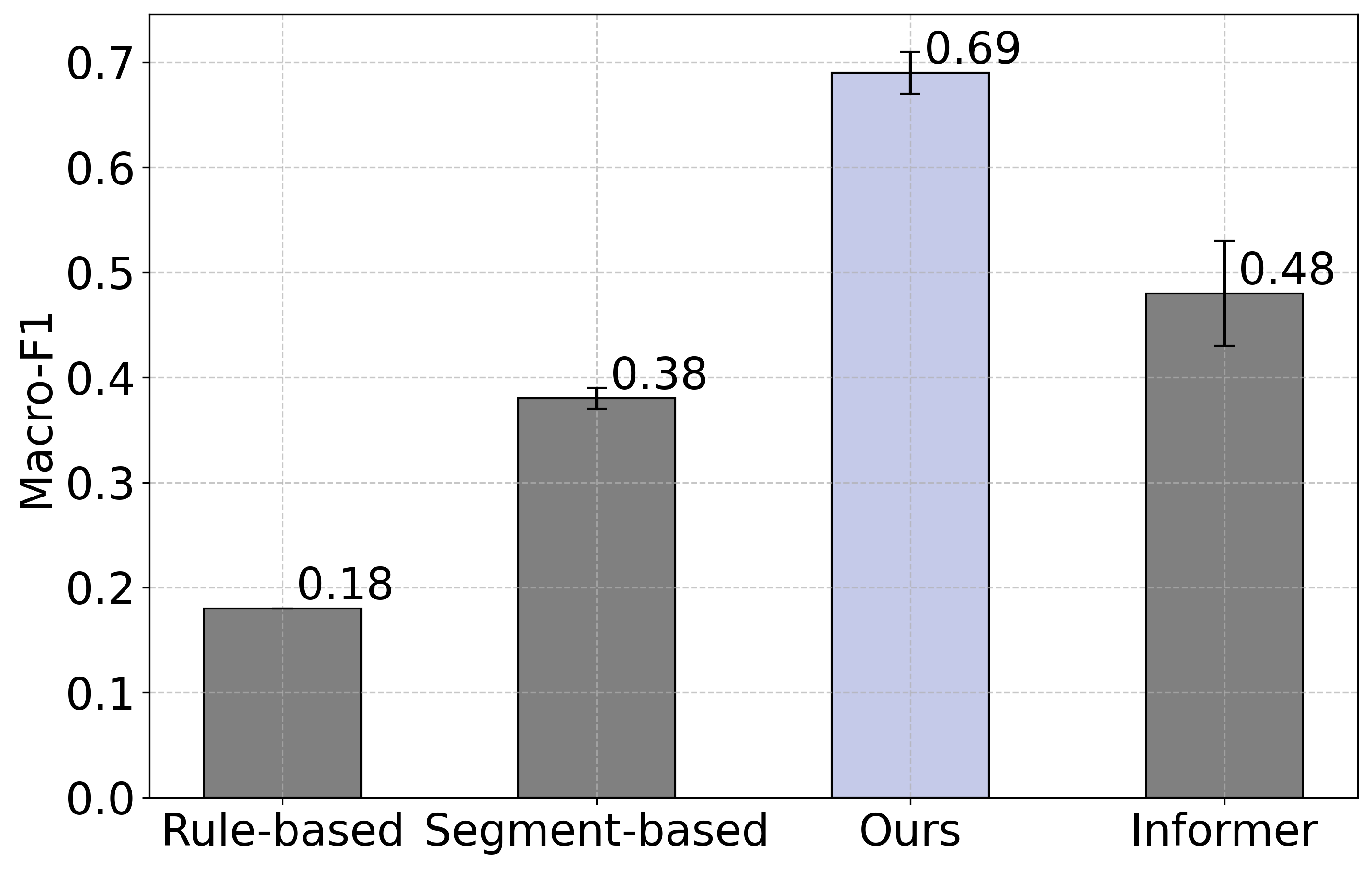}
            \captionsetup{font=small}
            \caption{\small Macro-F1 scores using abstract audio features like speaking status.}
            \label{fig:clf_all_feat_f1}
        \end{subfigure}
        \hspace{0.02\textwidth} 
        \begin{subfigure}[b]{0.48\textwidth}
            \centering
            \includegraphics[width=\linewidth]{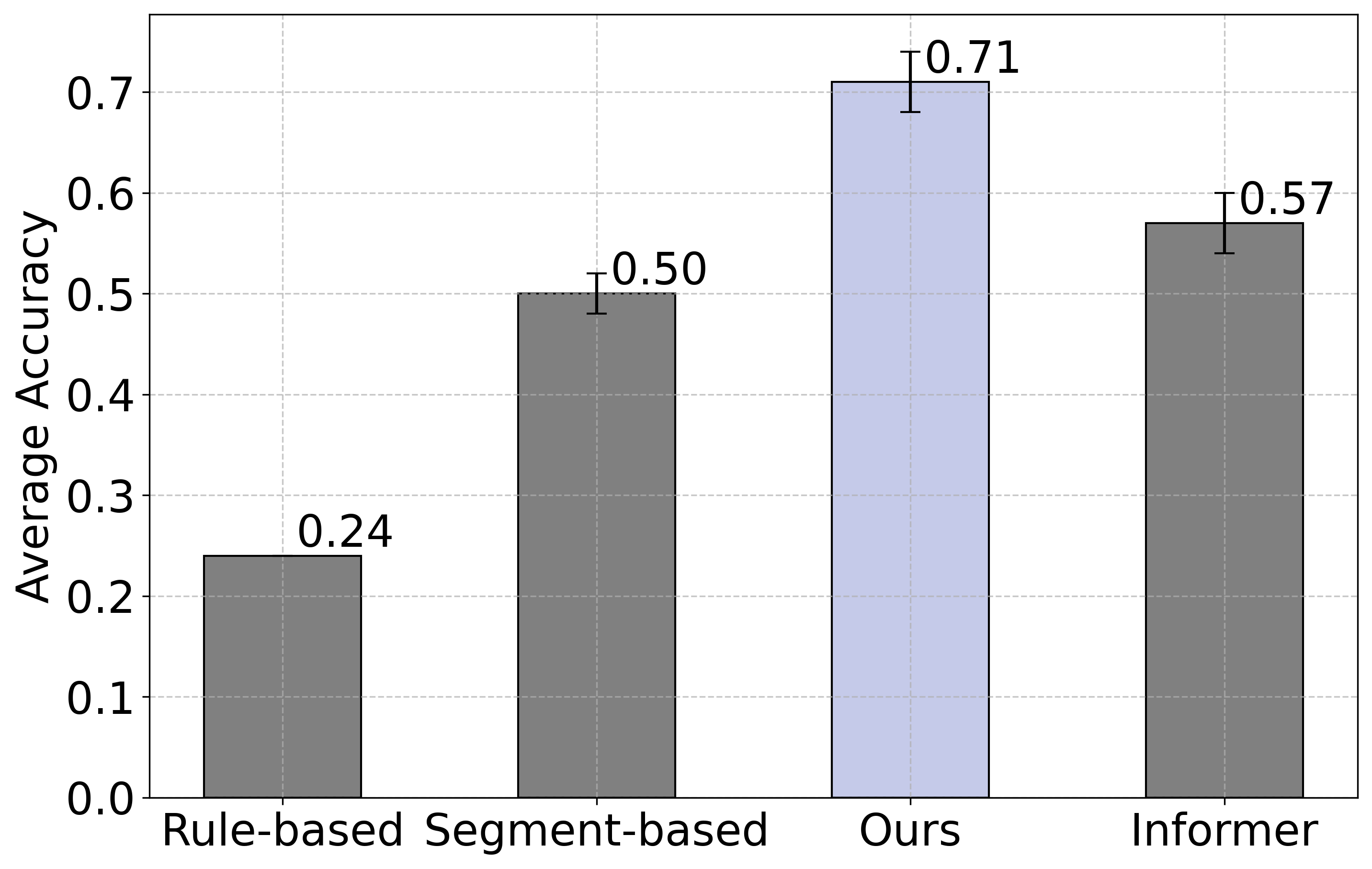}
            \captionsetup{font=small}
            \caption{\small Average accuracy using abstract audio features like speaking status.}
            \label{fig:clf_all_feat_acc}
        \end{subfigure}
        
        \vspace{0.5em} 
        
        \begin{subfigure}[b]{0.48\textwidth}
            \centering
            \includegraphics[width=\linewidth]{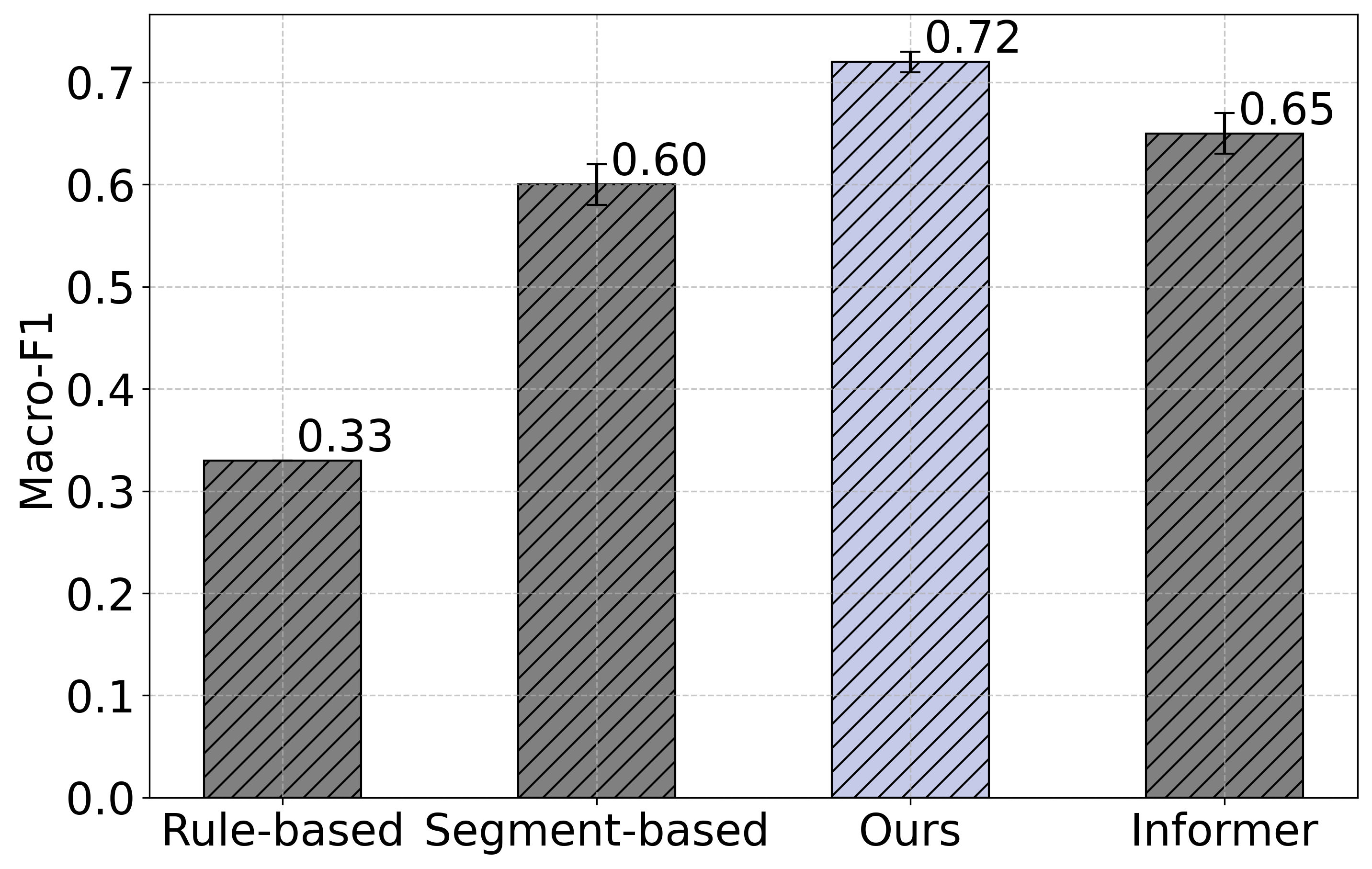}
            \captionsetup{font=small}
            \caption{\small Macro-F1 scores after target shaping.}
            \label{fig:clf_distil_f1}
        \end{subfigure}
        \hspace{0.02\textwidth}
        \begin{subfigure}[b]{0.48\textwidth}
            \centering
            \includegraphics[width=\linewidth]{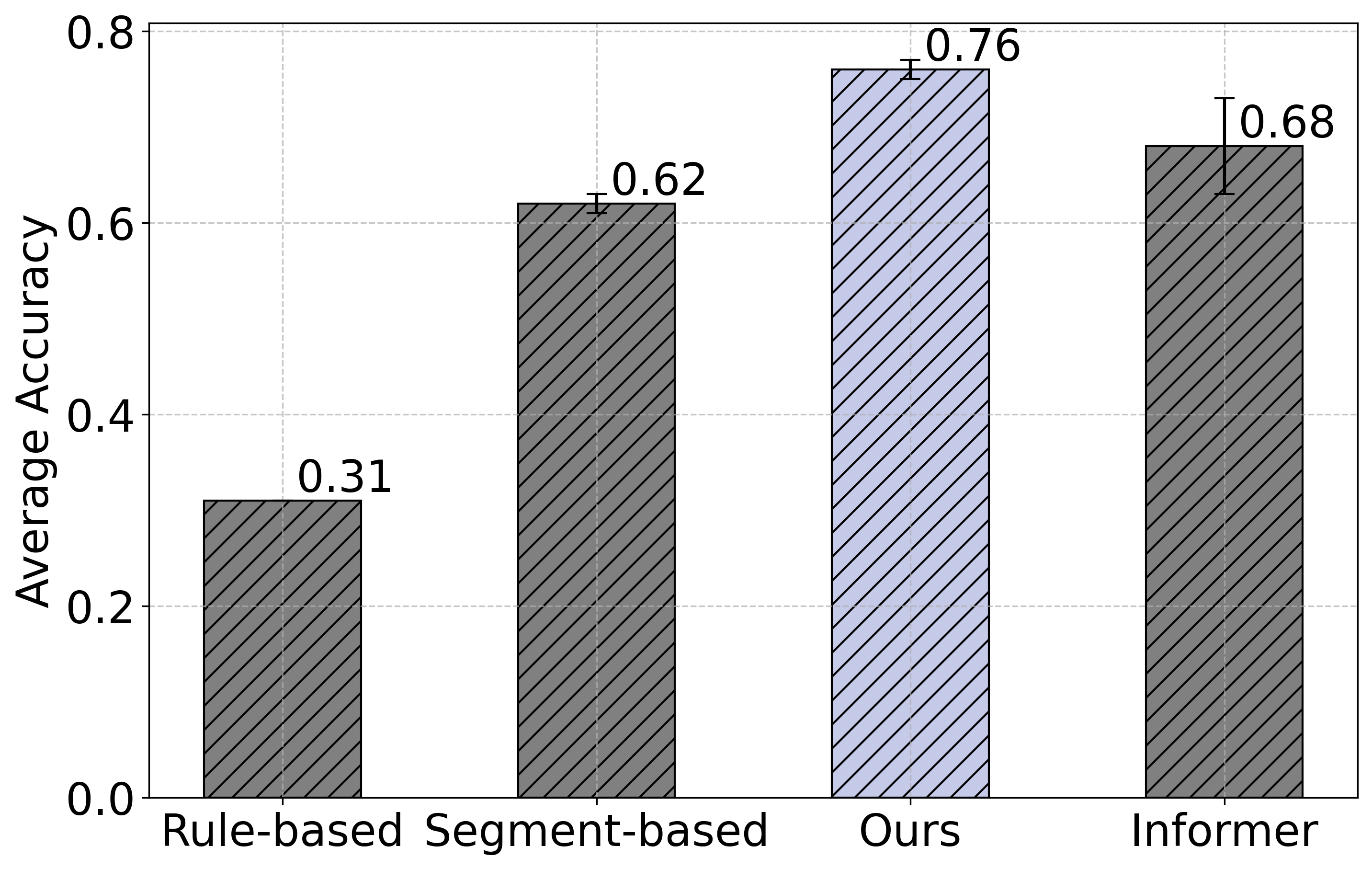}
            \captionsetup{font=small}
            \caption{\small Average accuracy after target shaping.}
            \label{fig:clf_distil_acc}
        \end{subfigure}
    \end{minipage}
    }
    
    \vspace{-8pt} 
    \captionsetup{font=small}
    \caption{\small Performance metrics for identifying the number of conversation partners. (a) and (b) use abstract audio features, while (c) and (d) reflect results after target shaping.}
    \label{fig:combined_results}
    \vspace{-15pt}
\end{figure}

\begin{wrapfigure}{r}{0.42\textwidth}
    \centering
    \includegraphics[width=0.95\linewidth]{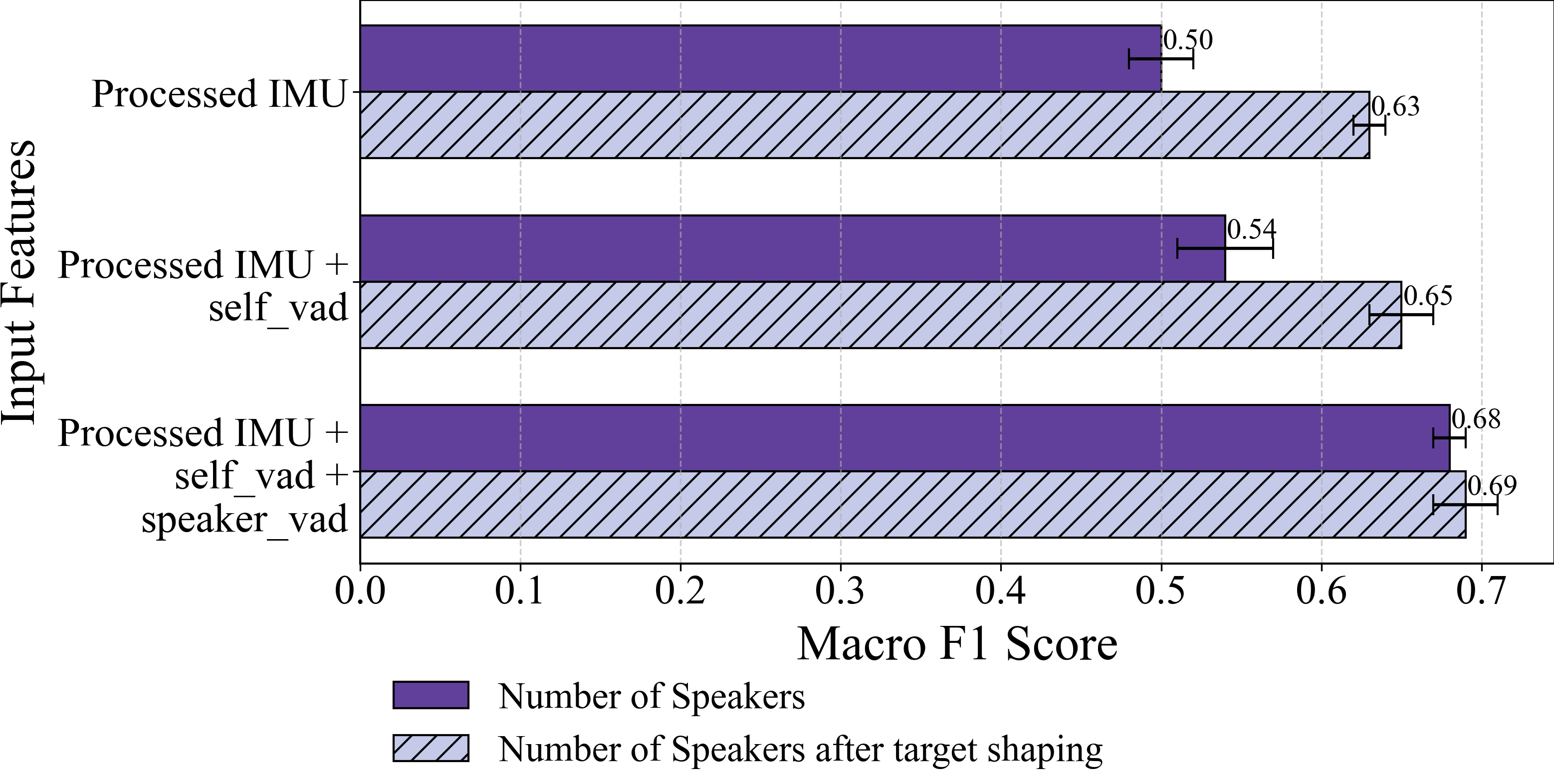}
    \vspace{-10pt}
    \caption{\imwut{Performance of classifying the number of conversation partners with additional abstract audio features and voice-activity-based target-shaping.}}
    \label{fig:clf_feat_importance}
    \vspace{-10pt}
\end{wrapfigure} \noindent \textbf{Abstract audio features and target shaping improve the identification of the number of conversation partners.} Figure~\ref{fig:clf_feat_importance} shows that utilizing only the information processed from IMUs achieves an average accuracy of 0.60 for four classes. Including the self-VAD and speaker-VAD boosts the accuracy to 0.73 or by applying target shaping based on the talkativeness of the conversation partners, as described in Section~\ref{sec:clf_design}, improvements up to 0.74, can be achieved. 

\noindent \textbf{\clfnet{} outperforms baselines in both the settings -- abstract audio features and target shaping.} Figures~\ref{fig:clf_all_feat_f1} and~\ref{fig:clf_all_feat_acc} compare the classification performance of all baselines using the enhanced feature set, while Figures~\ref{fig:clf_distil_f1} and~\ref{fig:clf_distil_acc} present the performance after target-shaping. While \clfnet{} demonstrates strong performance across these settings, it is noteworthy that target shaping significantly boosts the performance of the baselines. This highlights the promise of our simple, talkativeness-based statistical target-shaping approach and suggests the potential for incorporating more complex socio-auditory rules to design improved target representations.

Our results in Figures~\ref{fig:combined_results} and~\ref{fig:clf_feat_importance} illustrate the overall complexity of the classification task and highlights the benefits of either including abstract audio features, which are cheaper to obtain compared to a full audio-processing pipeline and intuitively help model more complex behavioral dynamics or incorporating domain-specific knowledge to qualify speakers as conversation partners \imwut{by target shaping as described in Section~\ref{sec:clf_design}}.


\subsection{Localization of Acoustic Zones of Interest with Relaxed Apriori Knowledge}\label{sec:clf_loc_comb}

Combining the designs for the localization and classification of conversation partners, we present results for \jointnet{}. To enable more effective joint optimization, we adopted the standard approach of stage-wise training \locnet{} and \clfnet{}, inspired by other domain-specific learning network architectures~\cite{gaddy-klein-2020-digital, mohapatra2023person}. Specifically, we extract the static features, \(\mathrm{d} \in \mathbb{R}^D\), from the penultimate layer of \(\mathrm{U}_{clf}\) of the trained \clfnet{} and fuse them with \(\mathrm{p}\) embeddings from the \locnet{}, resulting in \(\mathrm{r'} \in \mathbb{R}^{K+D}\). 

 Figure~\ref{fig:joint_opt_perf} compares the performance of the stand-alone \locnet{} with and without the actual number of conversation partners as static features, the performance when fusing noisy static features (by adding random uniform perturbations to static features in 30\% of the samples in each batch), and the performance using the estimated number of conversation partners representation from \clfnet{}. Our pilot results showed that a shared network (motivated by the overall similarity in design blocks for both tasks) did not perform well, achieving a Macro-F1 of 0.51. This is likely due to the complexity of the tasks, which require dedicated non-linearities to optimize each task objective. While larger datasets in the future may enable convergence to a general-purpose representation for extracting linguistic features from head orientation, we find that, in this case, stagewise training for both objectives performs better—almost on par with training using noisy static features. This suggests that incorporating a learned estimate of the static features representing the number of conversation partners may compensate for missing information. Although there is a 4\% decrease in performance compared to \locnet{} with an \textit{a priori} number of conversation partners, using \jointnet{} enables inference of acoustic zones of interest with minimal \textit{a priori} and additional features, which may otherwise be expensive to acquire in real-time.

\subsection{Additional Experiments for Design Validation}\label{sec:Opt_IMU_Perf}


\begin{figure}[!htbp]
    \begin{minipage}[t]{0.48\linewidth}
        \centering
        \includegraphics[width=\linewidth]{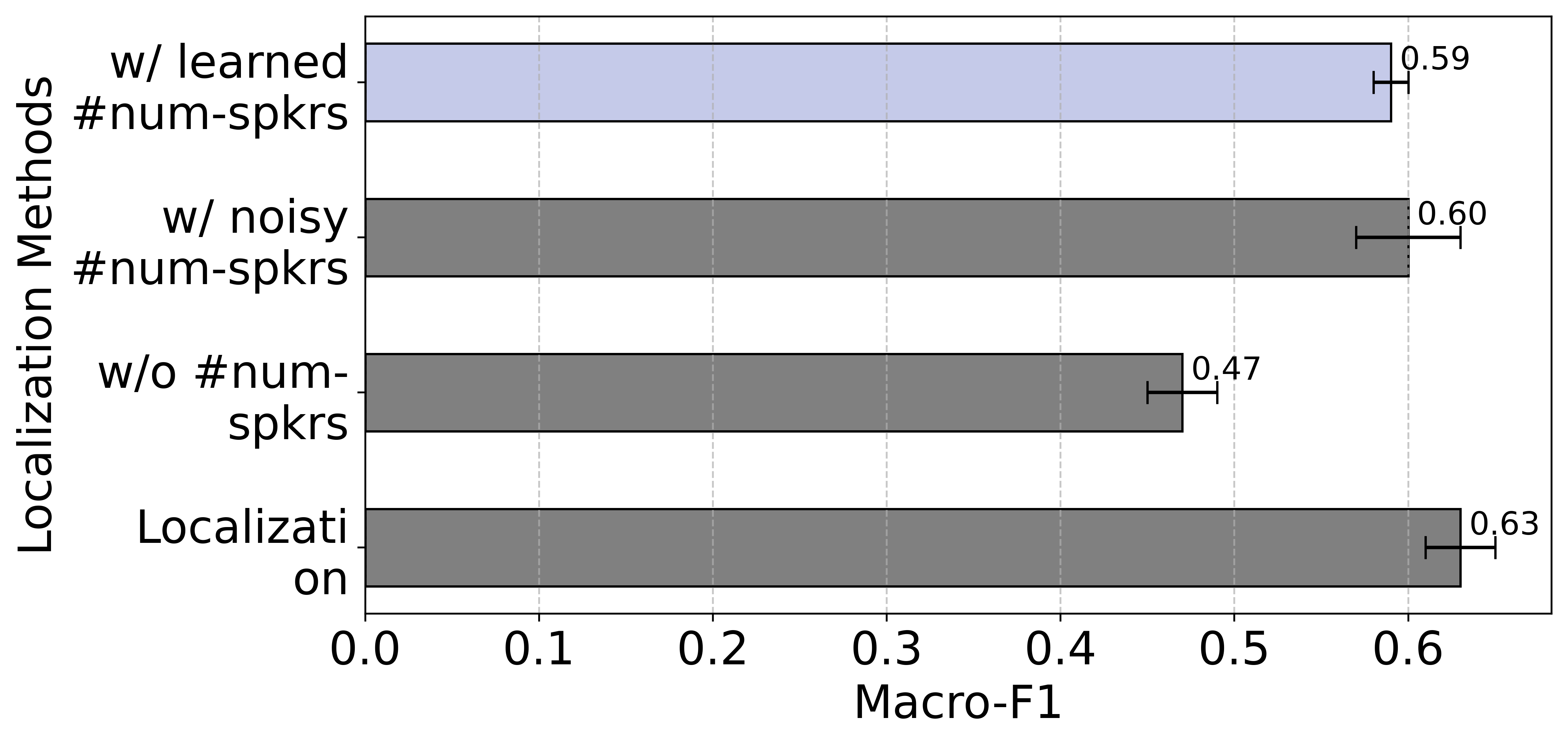}
        \captionsetup{font=small}
        \caption{\small Performance comparison of \locnet{} and its variants—without static features, with noisy static features, and using \jointnet{} with estimated static features.}
        \label{fig:joint_opt_perf}
    \end{minipage}%
    \hfill
    \begin{minipage}[t]{0.48\linewidth}
        \centering
        \includegraphics[width=0.8\linewidth]{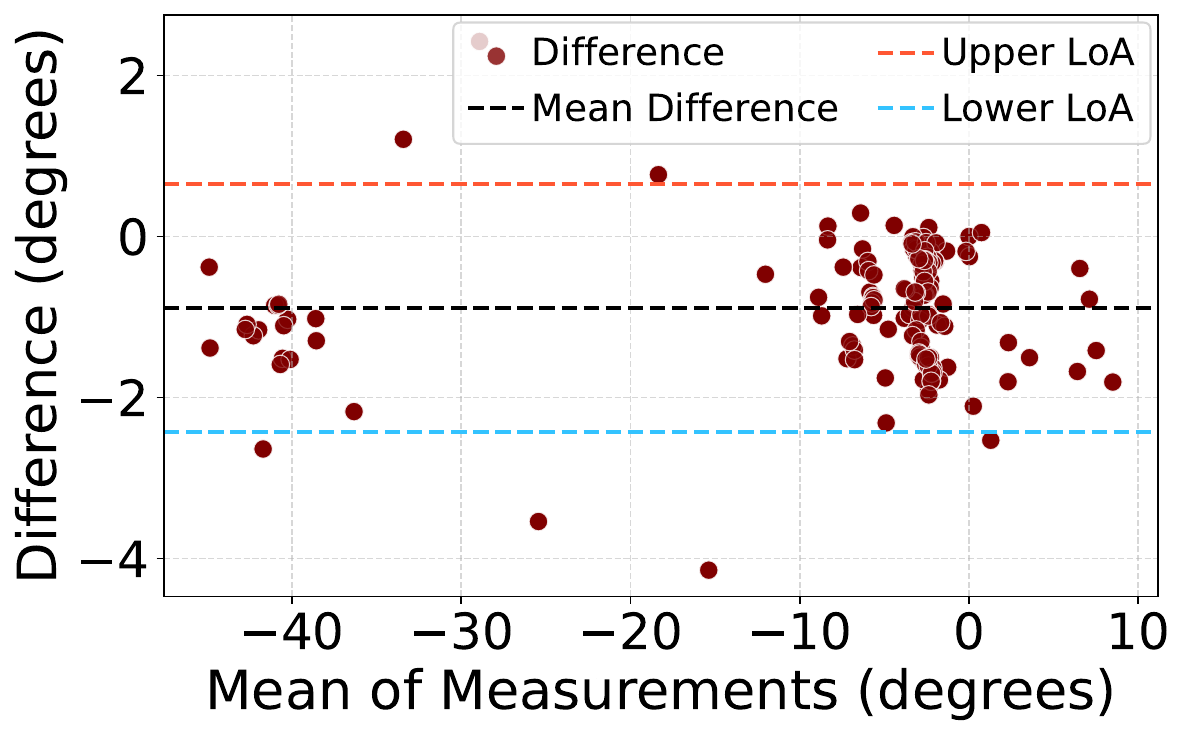}
        \captionsetup{font=small}
        \vspace{-10pt}
        \caption{\small Bland–Altman analysis for the complete dataset, showing a mean absolute error (MAE) of 0.97. 95\% of the samples lie within the limits of agreement for the overall 30-second segment.}
        \label{fig:bland_altman}
    \end{minipage}
    
\end{figure}



\noindent \textbf{Bland–Altman analysis of head orientation from IMUs vs. OptiTrack measurements.} We leverage the Bland–Altman statistical analysis tool~\cite{bland1986statistical}, commonly used to assess agreement between a new method and a gold-standard reference in clinical applications, to evaluate IMU-derived head orientation measurements against OptiTrack for the focal user. As shown in Figure~\ref{fig:bland_altman}, 95\% of the IMU-based head orientation measurements fall within $\pm$1.96 standard deviations of the differences compared to the ground truth from OptiTrack. This indicates strong agreement between the two methods for the chosen 30-second observation window. Our goal is to demonstrate the feasibility of extracting conversation-relevant features from short-duration IMU signals on smartglasses, without requiring explicit device-dependent sensor calibration.


\begin{figure}[!htbp]
    \centering
    \includegraphics[width=\textwidth]{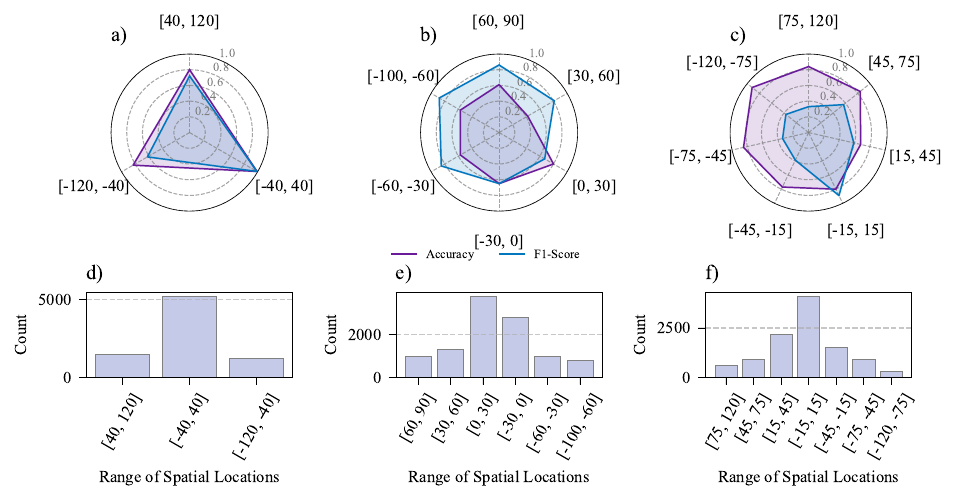}
    \captionsetup{font=small}
    \vspace{-20pt}
    \caption{\small Illustration of the F1-score and logit-wise accuracy for different spatial discretizations: (a) 3 spatial zones, (b) 6 spatial zones, and (c) 8 spatial zones, with their respective data distributions shown in (d-f), respectively. The convention followed in the paper is as follows: the frontal direction is 0$\degree$, the right is denoted by the negative angular direction, and the left by positive angular directions.}
    \vspace{-20pt}
    \label{fig:spat_loc}
\end{figure}
\noindent \textbf{Flexibility of azimuthal plane discretization.} The choice of spatial discretization for \locnet{} is application-specific. Figure~\ref{fig:spat_loc} reports the performance of three variants that differ only in the predictor head of \locnet{} (see Figure~\ref{fig:overal_arch}): (1) a simplified three-zone setup (front, left, right), (2) our original six-zone configuration, and (3) an eight-zone discretization. As expected, the three-zone discretization achieves the highest macro-F1 (0.78). Evident from Figures~\ref{fig:spat_loc}(d–f), with increasing discretization, the class imbalance at the extremities of the focal user’s field of view increases in the current dataset (refer Figures~\ref{fig:spat_loc}(a–c)), which leads to lower performance in the extremities of the spatial zones relative to the center. While we take steps to address these issues using standard techniques presented in Section~\ref{subsec:locnet_sec}, such as a weighted objective in the multilabel loss (Equation~\ref{eq:loc_loss}) and added non-linearity via an imbalanced classifier head, this also presents an opportunity to create more datasets with uniform seating layouts across the complete field of view of the focal user. Also, the higher performance of frontal direction may be attributed to the central bias in human head orientations during seated conversations~\cite{murdock2024self}.


\begin{wrapfigure}{r}{0.35\textwidth}
    \centering
    \includegraphics[width=\linewidth]{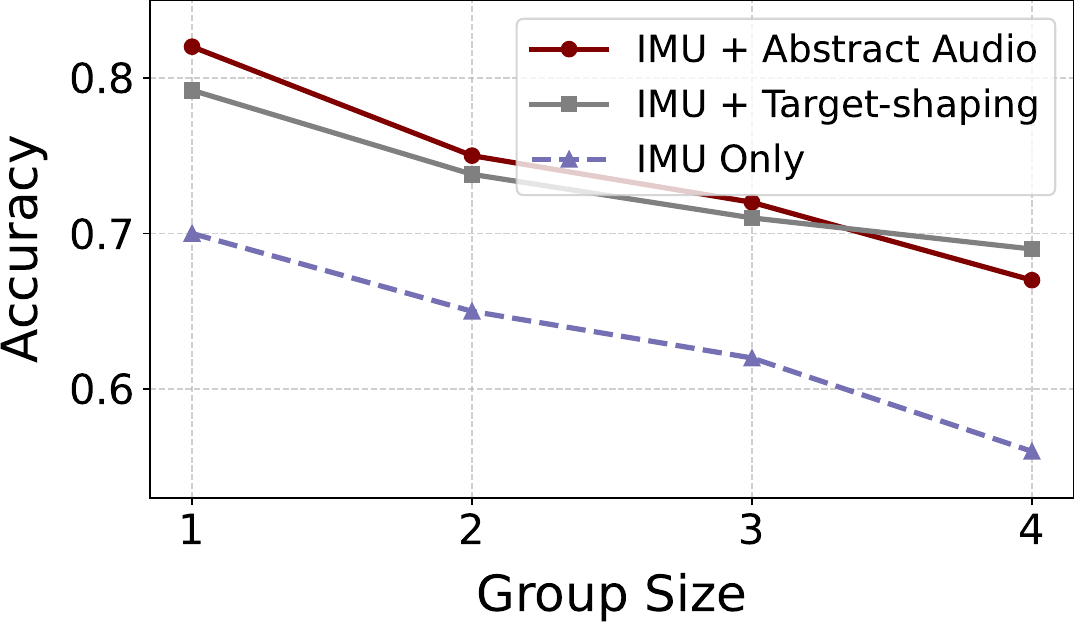}
    \vspace{-25pt}
    \captionsetup{font=small}
    \caption{\small \imwut{Comparison of IMU-only model with original targets versus proposed enhancements—(1) abstract audio features and (2) cumulative voice-activity-based target shaping—across increasing group sizes in multiparty conversations.}}
    \vspace{-10pt}
    \label{fig:group_acc_clf}
\end{wrapfigure} \noindent \textbf{Impact of increasing group size with different conditioning on \clfnet{}.} We analyze the performance of \clfnet{} with increasing group sizes, which overall exhibits a decreasing trend due to increased task complexity and fewer representative samples from the larger group sizes (> 3), as shown in Figure~\ref{fig:clf_stat}. However, Figure~\ref{fig:group_acc_clf} justifies the addition of abstract audio in the input features and voice-activity-based target shaping of the output, which boost the overall performance by an average of 17\% even under challenging conditions.

\imwut{
\noindent \textbf{Minimal end-to-end speech-enhancement pipeline using head orientation.} To demonstrate that head-orientation-based localization benefits real-world applications (Section~\ref{sec:related_cf} discusses these applications), we construct a minimal conversation-focusing system using a pair of smart glasses, which consists of a steering module that provides the steering direction to a beamformer and a spatial filter that enhances the signal in that direction\footnote{While exhaustive end-to-end augmented hearing systems are a substantial undertaking; our goal with this minimal study is to demonstrate head orientation’s practical utility as a localization front end for such systems using basic components.}. For this end-to-end study, we use the publicly available EasyCom dataset~\cite{donley2021easycom}, as it provides Array Transfer Functions (ATFs) for the microphones on the smart glasses, enabling speech‑enhancement algorithms to evaluate end‑to‑end performance. Although this dataset does not include IMU recordings, our earlier analysis (Figure~\ref{fig:bland_altman}) shows that IMU‑based features for short segments agree closely with OptiTrack‑based signals. Therefore, this dataset is suitable for use in our minimal system demonstration study. We closely follow the experimental setup of~\citep{donley2021easycom} and steer the highly directional beamformer using two modalities: (1) head-orientation-based: We use head orientation in two ways, first by using the frontal direction (where the user is looking) and second by using \locnet{}'s estimates of the zones of conversation partners; and (2) audio-based direction-of-arrival estimates using three common signal-processing algorithms—MUSIC~\cite{schmidt1986multiple}, GCC-PHAT~\cite{knapp2003generalized}, and SRP~\cite{do2007real}. We report the improved signal-to-noise ratio (SNR) values resulting from the enhanced speech using different steering methods. We present the implementation-specific details for this study in Section~\ref{app:e2estudy} of the Appendix and follow the standard definition of the performance metrics~\cite{donley2021easycom}.


\begin{wrapfigure}{r}{0.42\textwidth}
\centering
\begin{minipage}{\linewidth}
    \vspace{-8pt}

    \includegraphics[width=\linewidth]{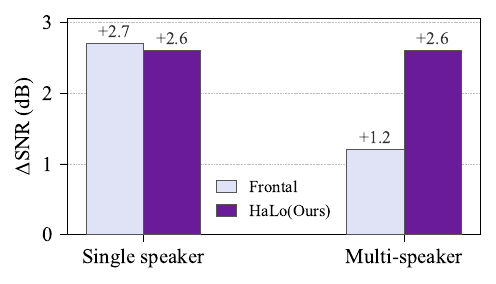}
    \captionsetup{font=small}
    \vspace{-25pt}
    \caption{\small \imwut{Our proposed world-locked partner localization using \locnet{} provides gains of +1.4 dB SNR in multiparty settings over the head-locked version, i.e., the user’s front-facing direction.}}
    \label{fig:frontal_halo}

    \includegraphics[width=\linewidth]{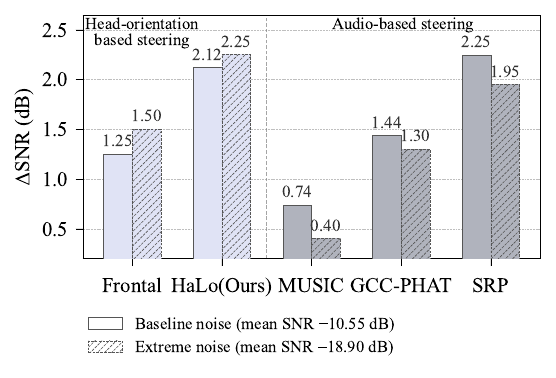}
    \captionsetup{font=small}
    \vspace{-25pt}
    \caption{\small \imwut{Comparing head-orientation (front facing and \locnet{}) and audio-based steering performance.}}
    \label{fig:snr_all}

\end{minipage}
\vspace{-10pt}
\end{wrapfigure}
The first observation is that the \textbf{front-facing direction of the focal user}, which is a head-locked version of steering the beamformer, \textbf{is not effective in multiparty conversations}, as shown in Figure~\ref{fig:frontal_halo}. This directly motivates our pursuit of a world-locked version of the system, where we can localize conversation partners irrespective of the front-facing direction of the focal user, which clearly offers greater benefit (by an absolute 1.4 dB). Additionally, it also paves the way for more naturalistic conversations, where the focal user is not tempted to proactively look in the direction of the conversation partner to enhance speech, and a world-locked beamformer can be steered based on the localization of the conversation partners.

The second observation is that, \textbf{in the current state, \locnet{}-based steering outperforms two out of three audio-based baselines by an average of $6\%$}, as shown in Figure~\ref{fig:snr_all}. The third observation is that, \textbf{interestingly, for extremely noisy samples} in the dataset, we can see that the \textbf{basic audio steering modules are not effective} (this behavior is also corroborated by previous works~\cite{qin2025robust, mitchell2023study, jiang2022egocentric}) \textbf{and that head-orientation-based steering offers a clear advantage}. This pattern likely arises from a fundamental difference between modalities: audio-based localization inevitably degrades as SNR decreases, whereas head-orientation signatures are unaffected by noise levels and may even become more informative. Under difficult listening conditions, the focal user may deliberately re‑orient toward conversation partners, producing stronger and more distinctive head‑orientation patterns that facilitate more accurate localization--features that \locnet{} is designed to exploit.

We present the implementation details and performance metrics in Table~\ref{tab:snr_metrics} in the Appendix. These results positively indicate that fusing head-orientation information with audio has the potential to offer superior performance in conversation-focusing applications.


}

%% file: sections/discussion.tex
\section{Scope and Limitations}


Our work studies a large dataset comprising over 71 unique sessions (yielding 7,915 samples), RLR-Chat, one of the few large-scale studies incorporating IMU data in natural, unscripted conversational contexts. However, due to the data collection timeline during 2020–2021, amid the pandemic, only family members participated in the study. \imwut{We focus on seated conversational settings to balance feasibility and practicality. As a result, the scope of our evaluation is limited in terms of the diversity of social configurations, seating geometries, and room layouts considered, and does not capture conversational dynamics in fully mobile or more socially complex environments.}


\imwut{In this work, we assume that the conversation partners remain fixed during each segment (in this case, 30s in duration). Supporting fully dynamic conditions, such as people leaving or joining, is a practical but very difficult problem, and that is why most previous works on speaker localization using audio~\cite{schmidt1986multiple, knapp2003generalized, do2007real} or audio-vision~\cite{jiang2022egocentric, murdock2024self, ryan2023egocentric} also investigate fixed conversation group settings. Recent audio-based works consider formalizing more pressing dynamic cases such as microphone-array movement, i.e., the focal user’s head motion~\cite{mitchell2023study}, and noise and reverberation conditions~\cite{mitchell2025improved, dong2019classification}. Future efforts can construct large-scale datasets with more diverse and unconstrained scenarios and incorporate behavioral and audio multimodal cues to handle such settings suitably.}

While we approached the task of identifying conversation partners as a classification problem using enhanced feature sets or targets, future work could explore advanced strategies, such as ordinal classification frameworks that penalize predictions based on the degree of error.

We will investigate other potential areas, such as inferring speaking states from head movements. Prior works, such as those by~\citet{tiede2019noggin} and~\citet{mcclave2000linguistic}, indicate that these states manifest in distinct head-orienting behaviors. Although these studies do not directly address voice activity detection, they demonstrate a strong coupling between head orientation and linguistic behavior like turn-taking and speaking status. Manual analyses of group conversations in these works provide compelling evidence for this relationship, which could further inform our research especially \clfnet{} network.

In this work, we demonstrate that head orientation is a promising behavioral modality for inferring a user’s acoustic zones of interest. These findings lay the foundation for the non-trivial, yet promising, integration of such behavioral cues into audio and multisensory processing pipelines. This integration can advance immersive conversation enhancement and hearing aid applications with smartglasses, particularly under challenging acoustic conditions.

%% file: sections/conclusion.tex
\section{Conclusion}
To develop practical conversation-enhancing technologies for wearable platforms like smart glasses, it is essential to understand users' preferences for acoustic zones of interest. While several neuroscience studies have highlighted the importance of behavioral modalities such as head orientation for understanding conversational components in controlled settings with small datasets, and deep learning research on large multimodal datasets—particularly audio-visual modalities—has emphasized the value of visual behavioral information in identifying users' preferred zones of auditory interest, our work explores a more challenging setting: leveraging only behavioral data from Inertial Measurement Units (IMUs) on smart glasses to infer acoustic zones of interest. We formulate and study two practical tasks on a large-scale dataset with minimal \textit{a priori} information: (1) localizing acoustic zones of interest, achieving an average accuracy of 0.78 and a macro-F1 score of 0.62 for multi-label classification across six discrete zones, and (2) determining the number of conversation partners, achieving an average accuracy of 0.74 in identifying group sizes ranging from 1 to 4 conversation partners. Our results highlight the importance of designing dedicated features, learning architectures, and objective functions, rather than relying on domain-specific statistical rule-based methods or general-purpose time-series representation learning approaches. Additionally, we provide in-depth qualitative analyses of the model's interpretability, its predictions over extended conversational settings, \imwut{an end-to-end speech enhancement pipeline that clearly supports the complementary benefits of a behavioral modality like head orientation}, and the role of various features and targets in performance. We present a promising direction for leveraging on-device IMU sensors on smart glasses to learn users' acoustic zones of interest based on head orientation and to inform advanced audio-enhancement solutions on wearables.

%% file: sections/appendix.tex
\newpage
\section*{APPENDIX}

\section{Additional Details on Data Preparation}

\subsection{Dataset Organization}
\begin{figure}
    \centering
    \includegraphics[width=0.7\linewidth]{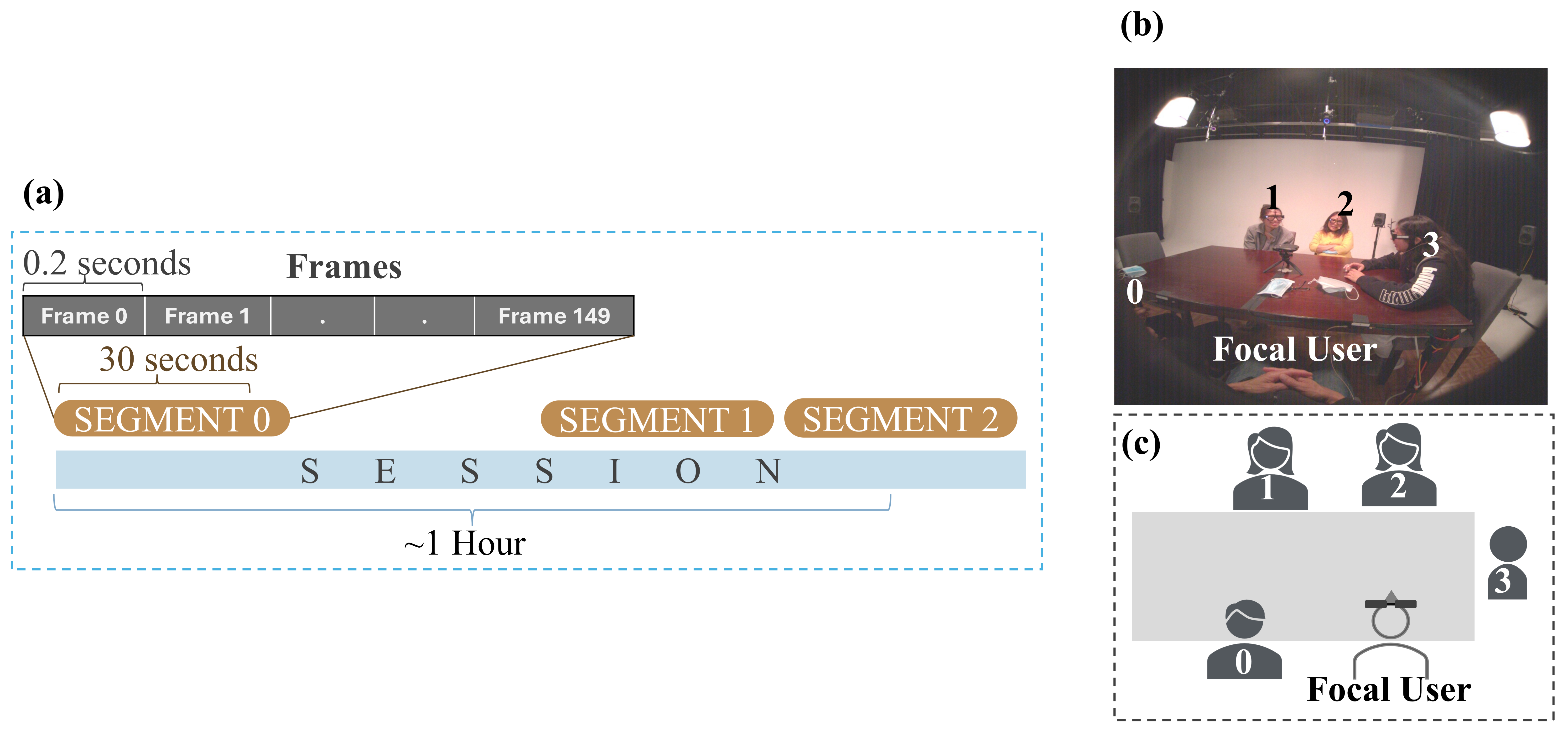}
    \caption{(a) Represents the RLR-Chat dataset organization where \textit{sessions} are manually validated for ordered but non-consecutive 30-second \textit{segments}, each consisting of 150 \textit{frames} for a group of participants. (b) illustrates the focal-user's viewpoint and (c) the exocentric viewpoint for a frame.}
    \label{fig:dataset_viz}
\end{figure}

The overall dataset is organized as shown in Figure~\ref{fig:dataset_viz}.

\subsection{Implementation Details of Head Orientation Approximation from IMUs} \label{app:imu_ho_geometry}
Our primary behavior-modality of exploration is the 6-axis IMU data collected by sensors on smart glasses. As noted by prior works, leveraging the translation from IMUs through standard double integration is prone to errors~\cite{yan2018ridi}, and many works propose using physics-guided learnable modules to determine global or relative translational motion~\cite{xu2024mobileposer}. However, in this case, we analyze seated natural conversations for auditory zones of interest, where translational motion does not offer much benefit. Since our task here is not tightly coupled with accurate head-pose estimation, instead, it focuses on the overall dynamics of head orientation, as evidenced by past neuro-speech studies. Consequently, we determine the approximate head orientation using a simple attitude integration scheme.

The IMU is placed on the legs of the smart glasses, as shown in Figure~\ref{fig:HO_spat_b}, and is used as a proxy for determining head rotation. The continuous angular velocity $\boldsymbol{\omega}(t)$ is sampled at discrete regular intervals of $\Delta t$, where the instantaneous angular velocity is given as $\boldsymbol{\omega}(t_n) = [\omega_x \, \omega_y \, \omega_z]^\mathrm{T}$ at discrete time $t_n = n\Delta t$. To estimate the angular displacement of the head, we leverage a simple attitude integration scheme~\cite{jia2019quaternions}. 

Since the orthogonal matrix used to represent the pure rotation $\theta$ along one of the axes in a three-dimensional space has only four independent elements, a quaternion representation $\mathbf{q} \in \mathbb{R}^4$ is used for computational efficiency~\cite{sola2017quaternion}. Leveraging key results from~\citet{jia2019quaternions}, we can describe the change in rotation $\Delta \mathbf{q}$ about the instantaneous axis, $\mathbf{u} = \frac{\boldsymbol{\omega}}{\|\boldsymbol{\omega}\|}$ during $\Delta t$ time units in the IMU's local time frame (rad/seconds in this case) as:

\[
\Delta \mathbf{q} = \cos \frac{\theta}{2} + \mathbf{u} \sin \frac{\theta}{2} = \cos \frac{\|\boldsymbol{\omega}\| \Delta t}{2} + \frac{\boldsymbol{\omega}}{\|\boldsymbol{\omega}\|} \sin \frac{\|\boldsymbol{\omega}\| \Delta t}{2},
\]
as a quaternion. We can consider the original state as $\mathbf{q}(t)$ rotated to a new state $\mathbf{q}(t + \Delta t) = \Delta \mathbf{q} \, \mathbf{q}(t)$~\cite{jia2019quaternions}. We have,

\begin{align*}
\mathbf{q}(t + \Delta t) - \mathbf{q}(t) &= \left( \cos \frac{\|\boldsymbol{\omega}\| \Delta t}{2} + \frac{\boldsymbol{\omega}}{\|\boldsymbol{\omega}\|} \sin \frac{\|\boldsymbol{\omega}\| \Delta t}{2} \right) \mathbf{q} - \mathbf{q} \\
&= \left( -2 \sin^2 \frac{\|\boldsymbol{\omega}\| \Delta t}{4} + \frac{\boldsymbol{\omega}}{\|\boldsymbol{\omega}\|} \sin \frac{\|\boldsymbol{\omega}\| \Delta t}{2} \right) \mathbf{q}
\end{align*}

Using the standard time-derivative property of quaternions~\cite{jia2019quaternions},
\begin{align*}
\dot{\mathbf{q}} &= \lim_{\Delta t \to 0} \frac{\mathbf{q}(t + \Delta t) - \mathbf{q}(t)}{\Delta t} \\
&= \lim_{\Delta t \to 0} \frac{1}{\Delta t} \left( -2 \sin^2 \frac{\|\boldsymbol{\omega}\| \Delta t}{4} + \frac{\boldsymbol{\omega}}{\|\boldsymbol{\omega}\|} \sin \frac{\|\boldsymbol{\omega}\| \Delta t}{2} \right) \mathbf{q} \\
&= \frac{\boldsymbol{\omega}}{\|\boldsymbol{\omega}\|} \lim_{\Delta t \to 0} \frac{1}{\Delta t} \sin \left( \frac{\|\boldsymbol{\omega}\| \Delta t}{2} \right) \mathbf{q} \\
&= \frac{\boldsymbol{\omega}}{\|\boldsymbol{\omega}\|} \frac{d}{dt} \sin \left( \frac{\|\boldsymbol{\omega}\| t}{2} \right) \bigg|_{t=0} \mathbf{q} \\
&= \frac{1}{2} \boldsymbol{\omega} \mathbf{q} \\
&= \frac{1}{2} 
\begin{bmatrix}
-\omega_x q_x - \omega_y q_y - \omega_z q_z \\
\omega_x q_w + \omega_y q_z - \omega_z q_y \\
\omega_y q_w + \omega_z q_x - \omega_x q_z \\
\omega_z q_w + \omega_x q_y - \omega_y q_x
\end{bmatrix}
\end{align*}

The definition of $\boldsymbol{\Omega}$ operator follows, 
\begin{equation}
\boldsymbol{\Omega}(\boldsymbol{\omega}) = 
\begin{bmatrix}
0 & -\boldsymbol{\omega}^T \\
\boldsymbol{\omega} & -[\boldsymbol{\omega}]_\times
\end{bmatrix} = 
\begin{bmatrix}
0 & -\omega_x & -\omega_y & -\omega_z \\
\omega_x & 0 & -\omega_z & \omega_y \\
\omega_y & \omega_z & 0 & -\omega_x \\
\omega_z & -\omega_y & \omega_x & 0
\end{bmatrix}
\label{eq:omega_init}
\end{equation}

We leverage~\citet{sola2017quaternion}'s results to arrive at the closed form solution for the new rotation, $\mathbf{q}_{t+1}$
\begin{equation} \label{eq:qt_1}
    \mathbf{q}_{t+1} = \left[ \cos \left( \frac{\|\boldsymbol{\omega}\| \Delta t}{2} \right) \mathbf{I}_4 + \frac{1}{\|\boldsymbol{\omega}\|} \sin \left( \frac{\|\boldsymbol{\omega}\| \Delta t}{2} \right) \boldsymbol{\Omega}(\boldsymbol{\omega}) \right] \mathbf{q}_t.
\end{equation}

The overall pseudocode to implement this is shown in Algorithm~\ref{alg:rot_update}. The device used for data collection was a pair of Aria glasses~\cite{engel2023project}, which experienced clock drift between devices. To mitigate this, the short-term audio's cross-correlation is used to fit a linear regressor and correct for the drift. Thus in this head-orientation extraction process, we also provide an opportunity to correct for any device-specific calibrations by converting the obtained $\mathbf{q}_{t+1}$ to the $\mathbb{R}^3$ coordinate space and cross product with $\mathbf{R}_{\text{ref}}$, a sensor-specific correction matrix. 

\begin{algorithm}[!htbp]
  \caption{Overview of Quaternion Update for Rotation Matrix Computation.}
  \label{alg:rot_update}
  \begin{algorithmic}[1]
    \State \textbf{Input:} $\mathbf{g}$: gyroscope readings (rad/s)
    \State \textbf{Input:} $\mathbf{q}_{\text{prev}}$: previous quaternion
    \State \textbf{Output:} $\mathbf{q}$: updated quaternion
    
    \Procedure{UpdateHO}{$\mathbf{g}, \mathbf{q}_{\text{prev}}, \Delta t$}
      \State Initialize $\boldsymbol{\Omega}$ from $\mathbf{g}$ following Equation~\ref{eq:omega_init}
      \State Compute angular velocity magnitude $w = \|\mathbf{g}\|$
      \State Calculate rotation matrix $\mathbf{A}$ from Equation~\ref{eq:qt_1}:
      \[
      \mathbf{A} = \cos\left(\frac{w \cdot \Delta t}{2}\right) \cdot \mathbf{I}_4 + \frac{\sin\left(\frac{w \cdot \Delta t}{2}\right)}{w} \cdot \boldsymbol{\Omega}
      \]
      \State Update quaternion $\mathbf{q} = \mathbf{A} \cdot \mathbf{q}_{\text{prev}}$
      \State Normalize quaternion $\mathbf{q} = \frac{\mathbf{q}}{\|\mathbf{q}\|}$
      \State \Return $\mathbf{q}$
    \EndProcedure
    
    \If{$\mathbf{q}_0$ is not provided}
      \State Initialize $\mathbf{q}_0 \gets [0, 0, 0, 1]$ 
    \EndIf
    
    \State Initialize quaternion array $\mathbf{Q} \gets \mathbf{0}_{N \times 4}$, where $N = \text{len}(\mathbf{g})$
    
    \For{each sample $n \in \{0, \ldots, N-1\}$}
      \If{$n = 0$}
        \State $\mathbf{q}_{\text{prev}} \gets \mathbf{q}_0$
      \Else
        \State $\mathbf{q}_{\text{prev}} \gets \mathbf{Q}[n-1, :]$
      \EndIf
      \State $\mathbf{Q}[n, :] \gets \textsc{UpdateHO}(\mathbf{g}[n, :], \mathbf{q}_{\text{prev}})$
    \EndFor
    
    \State Retrieve IMU calibration data to compute reference rotation $\mathbf{R}_{\text{ref}}$
    
    \For{each index $i \in \text{imu\_frame\_idx}$}
      \State Compute rotation matrix:
      \[
      \mathbf{R}_{\text{final}} = \mathbf{R}_{\text{ref}} \cdot \textsc{rotation\_from\_quat}(\mathbf{Q}[i, :]) \cdot \mathbf{R}_{\text{ref}}^{-1}
      \]
    \EndFor
    
    \State \Return $\mathbf{R}_{\text{final}}$ 
  \end{algorithmic}
\end{algorithm}

After computing the rotation matrix, $\mathbf{R}_{\text{final}}$, we transform a point, $\mathbf{v}_{\text{init\_xyz}}$, in the Cartesian plane to $\mathbf{v}_{\text{fin\_ae}}$ in the spherical coordinate system as shown in Algorithm~\ref{alg:head_rotation} using standard techniques~\cite{weisstein2005spherical} under some assumptions: (1) we do not model any translational motion and keep $d = 1$, (2) the front-facing direction of the participant wearing glasses is the origin, i.e., (azimuth, elevation) = $(0^\circ, 0^\circ)$.

\begin{algorithm}[!htbp]
  \caption{Overview of Head Rotation Computation.}
  \label{alg:head_rotation}
  \begin{algorithmic}[1]
    \State \textbf{Input:} \imwut{$\mathbf{v}_{\text{init}}$}: initial head orientation (azimuth, elevation) = $(0, 0)$
    \State \textbf{Input:} $d = 1$
    \State \textbf{Input:} $\mathbf{R}_{\text{final}}$: output from Algorithm~\ref{alg:rot_update} (or supplied externally)
    \State \textbf{Output:} $\mathbf{v}_{\text{fin\_ae}}$: final head orientation (azimuth, elevation)
    
    \State \imwut{$\mathbf{v}_{\text{init\_xyz}} = \textsc{sph2cart}(\mathbf{v}_{\text{init}}, d)$}
    \State \imwut{$\mathbf{v}_{\text{fin\_xyz}} = \mathbf{R}_{\text{final}} \cdot \mathbf{v}_{\text{init\_xyz}}$}
    \State \imwut{$\mathbf{v}_{\text{fin\_ae}} = \textsc{cart2sph}(\mathbf{v}_{\text{fin\_xyz}})$}
  \end{algorithmic}
\end{algorithm}

\section{Definitions of Performance Metrics} \label{app:perf}

We use the following metrics for the localization of the conversation partners task. Consider the predicted vector, $\hat{y} \in \{0, 1\}^6$, and the actual vector, $y \in \{0, 1\}^6$.

\[
\text{Hamming Score} = \frac{1}{n} \sum_{i=1}^{n} \frac{|y_i \cap \hat{y_i}|}{|y_i \cup \hat{y_i}|}
\]

\[
\text{Logit-wise Accuracy} = \frac{1}{n} \sum_{i=1}^{n} \mathbb{1}(\hat{y_i} = y_i)
\]

where $\mathbb{1}$ is the indicator that the predicted class $\hat{y_i}$ is equal to the true label, $y_i$.
\[
\text{Logit-wise F1} = \frac{2 \times \text{Precision}_{\text{logit}} \times \text{Recall}_{\text{logit}}}{\text{Precision}_{\text{logit}} + \text{Recall}_{\text{logit}}}
\]

\[
\text{Macro F1} = \frac{1}{C} \sum_{i=1}^{C} \frac{2 \times \text{Precision}_i \times \text{Recall}_i}{\text{Precision}_i + \text{Recall}_i}, \text{where C=6}
\]

For the task of identifying the number of conversation partners, we leverage the accuracy metric, as

\[
\text{Accuracy} = \frac{\text{TP} + \text{TN}}{\text{TP} + \text{TN} + \text{FP} + \text{FN}}
\]

where:
- \(\text{TP}\) (True Positives) are instances correctly predicted as positive,
- \(\text{TN}\) (True Negatives) are instances correctly predicted as negative,
- \(\text{FP}\) (False Positives) are instances incorrectly predicted as positive,
- \(\text{FN}\) (False Negatives) are instances incorrectly predicted as negative.

\section{\imwut{Implementation Details of the Minimal Speech-Enhancement System in Section~\ref{sec:Opt_IMU_Perf}}} \label{app:e2estudy}

\imwut{
In this section, we describe the implementation details of the minimal end-to-end system discussed in Section~\ref{sec:related_cf}. We use the EasyCom dataset~\cite{donley2021easycom}, which provides multi-microphone recordings collected using a pair of AR smart glasses equipped with six microphones distributed around the frame. The primary reason for choosing EasyCom in this study is that it includes array transfer functions, enabling the construction of a fully functional end-to-end speech-enhancement pipeline. The dataset includes synchronized headset signals, OptiTrack-based head-pose data, and ground-truth speech labels for multi-party conversational scenes. Although EasyCom does not contain raw IMU measurements, since the IMU-based pose estimates closely approximate short-segment OptiTrack estimates (as validated in Figure~\ref{fig:bland_altman}), it is suitable for this analysis.

To compare orientation-based steering with audio-based steering, we implement three widely used direction-of-arrival (DoA) estimation methods—MUSIC~\cite{schmidt1986multiple}, GCC-PHAT~\cite{knapp2003generalized}, and SRP-PHAT~\cite{do2007real}. Each of these algorithms infers the dominant sound direction by exploiting inter-microphone phase differences, but they differ in how they model spatial correlation and suppress reverberation.

Table~\ref{tab:snr_metrics} summarizes the resulting speech enhancement performance in terms of intelligibility and SNR improvements. While the absolute SNR values remain modest—consistent with the findings in the original EasyCom paper—this primarily reflects the inherently noisy conditions and complex conversational dynamics in the dataset. Our goal in this minimal implementation is not maximizing absolute enhancement scores, but demonstrating a fair, interpretable comparison between audio-driven and orientation-driven steering strategies using simple baseline components. The results confirm that orientation-informed steering remains stable under acoustically challenging conditions and can serve as an effective augmentation to conventional audio-based approaches for robust multimodal speech enhancement.
}

\begin{table}[t]
\centering
\small
\caption{\imwut{Comparison of head-orientation-based and audio-based steering methods for speech enhancement.}}
\label{tab:snr_metrics}
\begin{tabular}{l|c|cc|ccc}
\hline
\imwut{\textbf{Metric}} 
& \imwut{\textbf{Raw}} 
& \multicolumn{2}{c|}{\imwut{\textbf{Head-orientation-based}}} 
& \multicolumn{3}{c}{\imwut{\textbf{Audio-based}}} \\
\cline{2-7}
& \imwut{\textbf{Reference Mic}} 
& \imwut{\textbf{Frontal (0$^\circ$)}} 
& \imwut{\textbf{Halo (Ours)}} 
& \imwut{\textbf{MUSIC}} 
& \imwut{\textbf{GCC-PHAT}} 
& \imwut{\textbf{SRP}} \\
\hline
\imwut{SNR (dB)}  
& \imwut{-10.55} 
& \imwut{-9.30} 
& \imwut{-8.63} 
& \imwut{-9.26} 
& \imwut{-9.11} 
& \imwut{-8.34} \\
\imwut{STOI}      
& \imwut{0.34}   
& \imwut{0.39}  
& \imwut{0.41}  
& \imwut{0.40}  
& \imwut{0.41}  
& \imwut{0.38}  \\
\imwut{PESQ}      
& \imwut{1.10}   
& \imwut{1.16}  
& \imwut{1.15}  
& \imwut{1.15}  
& \imwut{1.16}  
& \imwut{1.10}  \\
\hline
\end{tabular}
\end{table}